\documentclass[aps,prb,10pt,reprint,groupedaddress,superscriptaddress]{revtex4-2}

\usepackage[usenames,dvipsnames]{xcolor}
\usepackage{amssymb}
\usepackage{graphicx}
\usepackage{amsmath}
\usepackage{txfonts}
\usepackage[bookmarks=true,colorlinks,citecolor=blue,urlcolor=blue]{hyperref}
\usepackage{braket}
\usepackage{babel}
\usepackage{blindtext}
\usepackage[normalem]{ulem}
\newcommand{\cd}[1]{c^{\dagger}_{#1}}
\newcommand{\co}[1]{c^{}_{#1}}
\newcommand{\bd}[1]{b^{\dagger}_{#1}}
\newcommand{\bo}[1]{b^{}_{#1}}
\newcommand{\fd}[1]{f^{\dagger}_{#1}}
\newcommand{\fo}[1]{f^{}_{#1}}
\newcommand{\e}{\mathrm{e}}
\newcommand{\kvec}{\mathbf{k}}
\newcommand{\qvec}{\mathbf{q}}
\newcommand{\rvec}{\mathbf{r}}

\begin{document}

\newcommand{\OurTitle}{Single-hole spectra of Kitaev spin liquids:\\from dynamical Nagaoka ferromagnetism to spin-hole fractionalization}

\title{\OurTitle}

\newcommand{\TUM}{\affiliation{Technical University of Munich, TUM School of Natural Sciences, Physics Department, 85748 Garching, Germany}}
\newcommand{\MCQST}{\affiliation{Munich Center for Quantum Science and Technology (MCQST), Schellingstr. 4, 80799 M{\"u}nchen, Germany}}
\newcommand{\Imperial}{\affiliation{Blackett Laboratory, Imperial College London, London SW7 2AZ, United Kingdom}}

\author{Wilhelm Kadow} \TUM \MCQST
\author{Hui-Ke Jin} \TUM \MCQST
\author{Johannes Knolle} \TUM \MCQST \Imperial
\author{Michael Knap} \TUM \MCQST

\begin{abstract}
The dynamical response of a quantum spin liquid upon injecting a hole is a pertinent open question. 
In experiments, the hole spectral function, measured momentum-resolved in angle-resolved photoemission spectroscopy (ARPES) or locally in scanning tunneling microscopy (STM), can be used to identify spin liquid materials. In this study, we employ tensor network methods to simulate the time evolution of a single hole doped into the Kitaev spin-liquid ground state. Focusing on the gapped spin liquid phase, we reveal two fundamentally different scenarios. For ferromagnetic spin couplings, the spin liquid is highly susceptible to hole doping: a Nagaoka ferromagnet forms dynamically around the doped hole, even at weak coupling. By contrast, in the case of antiferromagnetic spin couplings, the hole spectrum demonstrates an intricate interplay between charge, spin, and flux degrees of freedom, best described by a parton mean-field ansatz of fractionalized holons and spinons. Moreover, we find a good agreement of our numerical results to the analytically solvable case of slow holes. Our results demonstrate that dynamical hole spectral functions provide rich information on the structure of fractionalized quantum spin liquids. 
\end{abstract}

\maketitle

\noindent Quantum spin liquids have long captivated interest as strongly-correlated systems due to their intriguing properties, such as long-range topological entanglement and fractionalized excitations~\cite{Savary2017,Balents10,Zhou2017,Knolle2019,Broholm2020}. Theoretical models and classifications have been developed to understand these exotic states~\cite{Wen1991,Wen2002}. The Kitaev spin liquid provides a unique route to access spin liquids because of its exact solvability~\cite{Kitaev2006}. However, more relevant descriptions of materials include additional spin interactions spoiling exact solutions~\cite{Jackeli2009, Rau2014, Banerjee2016, Takagi2019}. Thus, mean-field analyses or numerical studies are required to explore their effects on the stability of the quantum spin liquid phase or the formation of competing magnetic orders~\cite{Chaloupka2010, Schaffer2012, Iregui2014, Gohlke2017}.

Doping holes into the Kitaev spin liquid introduces an additional layer of complexity to the situation, enabling further fascinating phenomena. The model can still be solved exactly for slow holes by associating flux, fermion, and plaquette quantum numbers to the holes~\cite{Halasz2014, Halasz2016}. Moreover, previous studies analyzed the influence of a finite hole density with additional spin terms in the form of Heisenberg interactions within mean-field theories~\cite{You2012, Hyart2012, Okamoto2013, Scherer2014, Mei2012}. Remarkably, they found several superconducting states, including possible topological superconductivity.

Another pending problem is the dynamics of holes in the limit of sparse doping or even for a single hole inserted into the spin-liquid ground state. This question is of particular interest for experimental probes such as angle-resolved photoemission spectroscopy (ARPES) and scanning tunneling microscopy (STM), which directly measure the hole dynamics.
Multiple plausible scenarios arise for the single-hole response. One possibility involves the fractionalization of the hole into holon and spinon quasiparticles. If these quasiparticles are deconfined, the hole spectral function will exhibit broad features and the spinons will determine the shape at the lowest energies at strong coupling. In a one-dimensional system, this scenario occurs naturally \cite{Meden1992,Voit1993, Giamarchi2004}, but is also relevant for spectral functions in higher dimensions~\cite{Senthil2008, Podolsky2009}. This was directly confirmed numerically for the kagome spin liquid~\cite{Laeuchli2004} and the chiral spin liquid on the triangular lattice~\cite{Kadow2022}. On the other hand, for confined spinons and holons, the spectrum shows a sharp, possibly renormalized, dispersion at low energies, as observed for the square lattice antiferromagnet~\cite{Dagotto1990, Martinez1991, Auerbach1991, Beran1996, Laughlin1997, Brunner2000, Mishchenko2001, Bohrdt2020a, Bohrdt2020, Wrzosek2021}.

\begin{figure}[b!]
\centering
\includegraphics[trim={0cm 0cm 0cm 0cm},clip,width=0.95\linewidth]{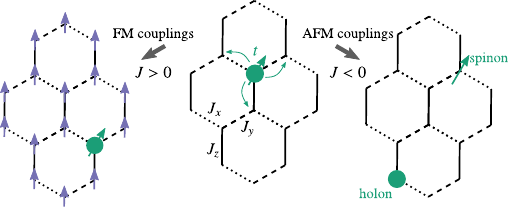}
\caption{\textbf{Response of Kitaev spin liquids upon single-hole doping.} When inserting a single hole into the Kitaev spin liquid, the hole either dynamically reorders the spin background into a Nagaoka ferromagnet for ferromagnetic (FM) spin couplings $J>0$ or fractionalizes into spinon and holon quasiparticles for antiferromagnetic (AFM) spin couplings $J<0$.}
\label{fig::schematic}
\end{figure}

Conversely, inserting a single hole can significantly impact the ground state of spin models. One famous example is the phenomenon of Nagaoka ferromagnetism~\cite{Nagaoka1966, Tasaki1989}, where the kinetic energy induced by the hole competes with the spin fluctuations. For strong interactions or equivalently fast holes, the ground state may then favor ferromagnetic order. First directly observed in quantum dot experiments~\cite{Dehollain2020}, this intriguing behavior has sparked recent explorations on frustrated lattices in the context of semiconductor heterostructures~\cite{Ciorciaro2023} and cold atoms~\cite{Xu2023, Lebrat2023, Prichard2023}.
However, this phenomenon does not only affect single-hole ground states but can also emerge dynamically, as has been demonstrated for states at infinite temperatures~\cite{Carlstroem2016, KanaszNagy2017}. So far, dynamical Nagaoka ferromagnetism from quenching a ground state with a hole has remained unexplored.

Considering these possible scenarios, this work investigates the crucial question of what happens to a quantum spin liquid state when a single hole is dynamically inserted. Specifically, we focus on the hole dynamics in the Kitaev spin liquid with both ferromagnetic and antiferromagnetic couplings; Fig.~\ref{fig::schematic}. As exact solvability is lost, we employ tensor network methods to simulate the time evolution after a hole quench and compute the hole spectral functions. In contrast to previous ARPES studies and high-temperature experiments of the Kitaev candidate materials \mbox{$\alpha$-RuCl$_3$~\cite{Zhou2016, Sinn2016}} and \mbox{Na$_2$IrO$_3$~\cite{Comin2012, Alidoust2016, Wang2023}}, we directly analyze the fate of the Kitaev spin-liquid ground state upon sudden hole injection.
Specifically, we show that for ferromagnetic Kitaev couplings, even slow-moving holes can dynamically polarize the ground state from a Kitaev spin liquid to a ferromagnetic state due to the dynamical Nagaoka effect. Intriguingly, for antiferromagnetic Kitaev couplings, the hole spectral function shows signatures of a spinon-holon fractionalization that a parton mean-field theory can partly explain. However, due to the complex interplay of spin, charge, and flux physics that renormalizes the energy scales of each other, the parton mean-field ansatz cannot capture all phenomena correctly. By contrast, in the limit of slow holes, we can directly confirm the contribution of fractionalized excitations~\cite{Halasz2014}.\\

\noindent \textbf{\large Results}

\noindent \textbf{Model}

\noindent 
The exactly solvable Kitaev model~\cite{Kitaev2006} is a paradigmatic example of a quantum spin liquid hosting topological order and fractionalized excitations. Here, we summarize some of its most important properties. The Kitaev model is given by
\begin{equation}\label{eq::Kitaev}
    H_K = -J_x \sum_{\langle i,j \rangle_x} \sigma^x_i \sigma^x_j
 -J_y \sum_{\langle i,j \rangle_y} \sigma^y_i \sigma^y_j
 -J_z \sum_{\langle i,j \rangle_z} \sigma^z_i \sigma^z_j,
 \end{equation}
where the first, second, and third terms act on $x$-, $y$-, and $z$-bonds of a honeycomb lattice, respectively, see Fig.~\ref{fig::schematic}. Throughout this work we will fix $|J_x|=|J_y|\equiv|J|=1$. The system undergoes a phase transition from a gapped $\mathbb{Z}_2$ topological phase for $|J_z|>2$ to a gapless phase for $|J_z|<2$. Remarkably, these phases exist for both signs of the Kitaev couplings $J$. Here, we will examine both ferromagnetic couplings $J>0$ and antiferromagnetic couplings $J<0$ and focus on the gapped case. We find qualitatively similar behavior for the gapless phase; see ``Methods'' section for additional data.

\begin{figure}[t!]
\centering
\includegraphics[trim={0cm 0cm 0cm 0cm},clip,width=0.95\linewidth]{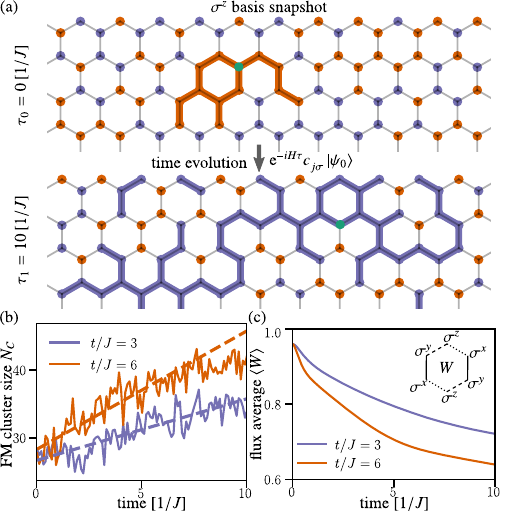}
\caption{\textbf{Dynamically emerging Nagaoka ferromagnetism in a doped Kitaev spin liquid with FM spin couplings.} (a) Typical examples for snapshots of the Fock configurations occurring with large probability in the wave function; see Eq.~\eqref{eq::coefficients}; obtained from matrix product states (MPS) at times $\tau=0\,[1/J]$ and $\tau=10\,[1/J]$ are shown. We identify the FM cluster size $N_C$ by connecting all spins pointing up (purple dots) or down (orange dots) starting from the hole site (green dots). (b) Extracted from 10,000 snapshots of the state at each time, a large ferromagnetic (FM) cluster of spins forms around the hole. Of the $N=160$ sites in the system, the cluster has a size of $N_c\approx 35\,(40)$ sites for hopping constants $t/J=3\,(6)$. The speed of the FM cluster expansions is set by the hopping constant $t$; see dashed lines fitted to the snapshot data. (c) While the FM cluster size increases, the flux operator average $\langle W \rangle$ decays with time. All data shown is for ferromagnetic couplings $J>0$ and anisotropy $J_z/J = 2.5$.}
\label{fig::nagaoka}
\end{figure}

The Kitaev model hosts an extensive number of conserved quantities, described by the flux operator around a plaquette; see inset in Fig.~\ref{fig::nagaoka}~(c):
\begin{equation}
    W = \sigma^x_1 \sigma^y_2 \sigma^z_3 \sigma^x_4 \sigma^y_5 \sigma^z_6
\end{equation}
For each plaquette $W$ has eigenvalues $\pm 1$. The ground state is in the \textit{flux free} sector, where $\langle W \rangle = +1$ for all plaquettes. Accordingly, we say that a flux is introduced into the system if $\langle W \rangle = -1$ for one of the plaquettes.

For the exact solution of Eq.~\eqref{eq::Kitaev}, the spin operators are decomposed into matter Majorana fermions $\chi^0_i$ and bond Majorana fermions $\chi^a_i$ [$a \in (x, y, z)$], such that $\{\chi^\alpha_i, \chi^{\alpha'}_j\} = 2\delta_{ij}\delta_{\alpha\alpha'}$ [$\alpha \in (0, x, y, z)$].
Then, the spins are written in terms of the Majorana fermions as
\begin{equation}
    \sigma_i^a = i\chi_i^0\chi_i^a, \quad a \in (x, y, z).
\end{equation}

Introducing directed bond operators $\hat{u}_{\langle i j \rangle_a} = i \chi^a_i \chi^a_j$ from the $A$ to the $B$ sublattice, we rewrite the Kitaev model
\begin{equation}\label{eq::Kitaev_majorana}
    H_K = i\sum_{a, \langle i j\rangle_a} J_a \hat{u}_{\langle i j \rangle_a} \chi^0_i \chi^0_j.
\end{equation}
The gauge operators $\hat u_{\langle i j \rangle_a}$ commute with each other and with the matter Majorana fermions. Thus, the Hamiltonian splits into separate sectors of the bond operators, which have eigenvalues $u_{\langle i j \rangle_a} = \pm 1$. Moreover, the plaquette operator $W$ can be expressed as a product of $\hat u_{\langle i j \rangle_a}$ around a hexagon as $W = \hat u_{\langle 1 2 \rangle_x}\hat u_{\langle 2 3 \rangle_y}\hat u_{\langle 3 4 \rangle_z}\hat u_{\langle 4 5 \rangle_x}\hat u_{\langle 5 6 \rangle_y}\hat u_{\langle 6 1 \rangle_z}$.

To introduce holes into the system, two different approaches have been discussed before. On the one hand, in Refs.~\cite{You2012, Hyart2012, Okamoto2013, Scherer2014, Trousselet2014} holes are inserted in a $t-J$ model formalism, similarly to the description of cuprates from a spin Hamiltonian on a square lattice~\cite{Damascelli2003}. On the other hand, Refs.~\cite{Halasz2014, Halasz2016} introduce holes as spin sites, where the interactions to all other sites are missing. While the second approach gives exact results for very slow holes, we will focus on the first one, which captures the experimentally relevant finite hopping regime. Later, we compare it to the slow-hole limit as well.
Therefore, we add a kinetic term for nearest-neighbor hole hopping to the Kitaev Hamiltonian:
\begin{align}\label{eq::Kitaev_tJ}
    H &= H_t+H_K \nonumber \\
    &= -t\sum_{\langle i j\rangle, \sigma}\mathcal{P}_\mathrm{GW}\left(\cd{i\sigma} \co{j\sigma} + \mathrm{h.c.} \right)\mathcal{P}_\mathrm{GW} - \sum_{a, \langle i j\rangle_a} J_a \sigma^a_i \sigma^a_j.
\end{align}
$\cd{j\sigma}$ ($\co{j\sigma}$) create (annihilate) fermions with spin $\sigma$ and the Gutzwiller projector $\mathcal{P}_\mathrm{GW}$ excludes doubly occupied sites. The holes are related to the spin operators by $\sigma^a_i=(\cd{i\uparrow}, \cd{i\downarrow}) \sigma^a (\co{i\uparrow}, \co{i\downarrow})^T$, where $\sigma^a$ are the corresponding Pauli matrices. The spin anisotropy in the Kitaev model originates from the strong spin-orbit coupling of electrons to the $d$-orbitals, for example, in Iridate materiales~\cite{Jackeli2009}. Therefore, the spins above are not electron spins but rather hybridized pseudo-spins. Nevertheless, they can directly be related to the electron operators at low energy as in Eq.~\eqref{eq::Kitaev_tJ}, which have an isotropic hopping; see derivation in the ``Methods'' section.

When including the term $H_t$, the Hamiltonian is no longer exactly solvable and also yields different behavior for AFM couplings $J<0$ and FM couplings $J>0$. Note that we stick to the usual Kitaev conventions with a minus sign in front of the spin Hamiltonian. Mean-field approaches offer convenient means to obtain results in this limit~\cite{You2012, Hyart2012, Okamoto2013, Scherer2014, Choi2018}; however, they possess inherent limitations and dependence on physical assumptions. Therefore, we will resort to numerical techniques to solve the dynamics of Eq.~\eqref{eq::Kitaev_tJ}. We note that exact diagonalization studies of a similar system have been performed earlier~\cite{Trousselet2014}. These are restricted to small system sizes and can hence only access a limited number of momenta for the hole spectral function.
Here we use tensor networks in the form of matrix product states (MPS), which allow us to study larger systems on a cylinder with length $L_x$ and circumference $L_y$. Details on the numerical methods are provided in the ``Methods'' section. \\
\begin{figure*}[t]
\centering
\includegraphics[trim={0cm 0cm 0cm 0cm},clip,width=1.0\linewidth]{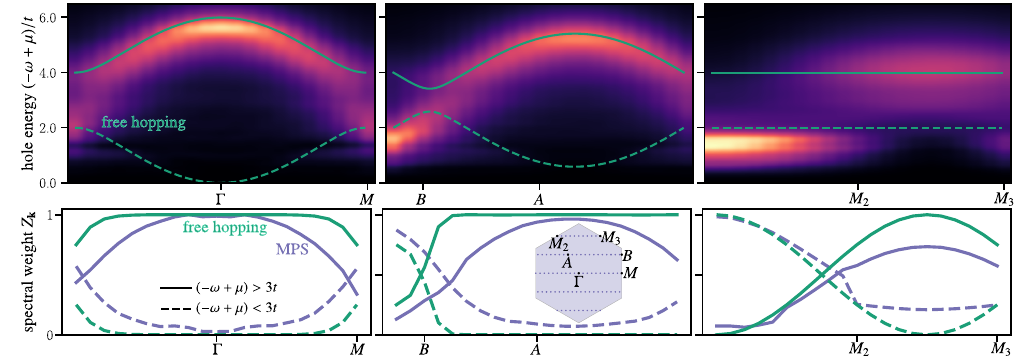}
\caption{\textbf{Hole spectral function for ferromagnetic Kitaev couplings.} The spectral function $A(\mathbf{k},\omega)$ for ferromagnetic $J>0$, $J_z/J = 2.5$ and $t/J=3.0$ along the three distinct cuts in the Brillouin zone (see inset in the bottom row) shows significant resemblance with the spectrum of a free hole hopping on a honeycomb lattice (green lines). Both the energy dispersion (top row) and the spectral weight (bottom row) agree well with the free hole.}
\label{fig::arpesFM}
\end{figure*}

\noindent \textbf{Dynamical Nagaoka ferromagnetism}

\noindent The presence of a ferromagnetic (FM) Kitaev coupling constant ($J > 0$) is expected to be realized in some prominent Kitaev materials, such as Na$_2$IrO$_3$~\cite{Choi2012} or \mbox{$\alpha$-RuCl$_3$~\cite{Winter2017}}. This motivates us to investigate the behavior of a single hole that is inserted into the spin-liquid ground state with FM couplings. 

We first look at snapshots of the state during the time evolution and analyze these to shed light on the underlying dynamics. With the perfect-sampling algorithm introduced in Ref.~\cite{Ferris2012}, we can extract typical product-state configurations of the MPS and the corresponding probabilities in the computational basis, which is either a spin-up, spin-down, or hole state for the $t$-$J$ model. For this basis $\{\ket{s} \equiv \ket{s_1s_2\dots}; s_i\in (\uparrow,\downarrow,\circ) \}$, the wave function is given by
\begin{equation}\label{eq::coefficients}
    \ket{\psi} = \sum_{s} c_{s} \ket{s}.
\end{equation}
Then the probability for a snapshot (i.e., the Fock space configuration $|s\rangle$) is $p(s)=|c_{s}|^2$. We can use this for sampling the expectation value of an operator $O$ with a subset $\mathcal{S}$ of basis states, where $O$ is diagonal in the corresponding basis,
\begin{equation}\label{eq::sampling}
    \bra{\psi} O \ket{\psi} \approx \sum_{s \in \mathcal{S}} p(s) \bra{s} O \ket{s}.
\end{equation}

Note that particle number conservation enforces precisely one hole per snapshot. Examples of such snapshots are displayed in Fig.~\ref{fig::nagaoka}~(a). For the initial state at $\tau=0\,[1/J]$, some small FM clusters are found, which are formed only from the spin background of the Kitaev spin liquid due to quantum fluctuations. At late times, e.g., $\tau=10\,[1/J]$, we observe larger FM clusters around the site where the hole is located. The anisotropic spin interactions $J_z/J = 2.5$ used in our simulations favor FM correlations in $z$-basis, which is used as the Fock basis to sample the snapshots. Hence, the snapshots directly depict the fluctuating FM order.

The system undergoes a complex evolution over time. We define the FM cluster size $N_c$ around a hole as the number of aligned spins that directly connect to the site where the hole is for a given snapshot. Note that for the initial state, the hole will always be at the origin, where it was inserted. However, after some time, it will have spread to different lattice sites. Therefore, we cannot capture the emergent cluster size with (local) expectation values but have to resort to the analysis of snapshots. The average of 10,000 snapshots reveals that the cluster size increases linearly in time; see Fig.~\ref{fig::nagaoka}~(b). The velocity of this increase depends on the hopping parameter~$t$, as demonstrated by the presented fits (dashed lines). Notably, for fixed system size, the maximum size of the FM clusters is proportional to the ground state magnetization $M_0=\sum_i \langle \psi_{0,\text{1h}} |\sigma_i^z |\psi_{0,\text{1h}}\rangle$ in the presence of one hole, which also is approximately proportional to $t$~\cite{Hui-Ke2023}. However, due to the energy constraints within unitary time evolution, the FM clusters cannot reach the ground-state magnetization value $M_0$ as the hole excites the system. Instead, we find $N_C \sim M_0/2$, with $M_0$ determined numerically. We also find that increasing $t$ or doping a finite but small hole density instead of a single hole leads to the same magnetization for larger system sizes.

The dynamical formation of the FM cluster can be understood from the same argument as in Nagaoka's theorem for ground states with a single hole. For a hole that hops on a bipartite lattice, the kinetic energy is minimized when the interference of different hopping paths is constructive. In particular, this is the case when all spins are aligned, and hence the hole motion does not change the spin pattern. The minimization of kinetic energy usually competes with antiferromagnetic spin interactions, which prefer anti-aligned spins. For example, for the square lattice AFM, a large hopping strength is required to establish Nagaoka ferromagnetism. There, numerical computations find a large FM polarization around the hole for $t_\mathrm{Nagaoka}/J \gtrsim 30$~\cite{White2001}. In contrast, for the Kitaev model, the local spin interaction is already ferromagnetic and frustration only arises between the different $x-$, $y-$, and $z-$ spin directions. Thus, a small hole hopping $t/J$ and anisotropy $J_z$ can easily favor the ferromagnetism through kinetic energy minimization.
This is the reason for the formation of large FM clusters around the hole, already for the moderate hopping strength of $t/J=3(6)$ considered in Fig.~\ref{fig::nagaoka}.

The flux average $\langle W\rangle$ over all plaquettes of the system, Fig.~\ref{fig::nagaoka}~(c), quickly decreases with time which indicates the disappearance of the Kitaev spin liquid. This observation highlights the transformative nature of the system as it transitions from a spin liquid to a ferromagnetic state. However, finite flux $\langle W\rangle$ can still be present simultaneously with large FM clusters, which is a remnant of the Kitaev spin liquid. Even though the spins get dynamically polarized, information remains about the initial flux configuration. The same phenomenon occurs in the ground state of the one-hole doped Kitaev model, where both the magnetization and the flux expectation value $\langle W\rangle$ can be non-zero simultaneously~\cite{Hui-Ke2023}.

The hole spectral function, as measured by ARPES, provides energy- and momentum-resolved information about the response of the spin-liquid state. It gives insights into the fractionalization of quasiparticles in terms of spinons and holons~\cite{Laeuchli2004, Podolsky2009, Bohrdt2018, Kadow2022}.
Concretely, we are interested in calculating
\begin{equation}
    A(\mathbf{k},\omega) = \sum_{\sigma=\uparrow,\downarrow} \int \mathrm{d} \tau \e^{i\tau \omega} \bra{\psi_0}\cd{\mathbf{k}\sigma}(\tau) \co{\mathbf{k}\sigma}(0) \ket{\psi_0},
\label{eq::arpes}
\end{equation}
where $\ket{\psi_0}$ denotes the Kitaev spin-liquid ground state. For numerical details and additional data regarding convergence in bond dimension, we refer to the ``Methods'' section. We shift all energies by a constant $\mu$ that is computed from the energy difference between the ground state without a hole and with a single hole. Therefore, it corresponds to the Mott gap of the insulator, and all hole excitations have to be outside that gap. In our convention, we plot $-\omega+\mu$, which hence must be a positive energy.

In previous X-ray scattering studies~\cite{HwanChun2015}, the bond-depending interactions could be directly observed by measuring different spin components separately. Similarly, a spin-resolved response of the hole spectrum is experimentally feasible~\cite{Lin2021}. We derive the connection between the spectrum for all physical electrons in a d$^5$ orbital and the low-energy effective electrons as in Eq.~\eqref{eq::arpes} in the ``Methods'' section. However, in our numerical computations, we observe the same responses for spin up and down for all spectral functions below, corresponding to the symmetry of $\cd{j\uparrow}$ and $\cd{j\downarrow}$ in Eq.~\eqref{eq::Kitaev_tJ}. Therefore, we cannot gain additional insights from looking at the spin-resolved spectra numerically. However, spin-asymmetry may be used in experiments to characterize perturbations to the considered model. 

The resulting spectrum for $J>0$ with a Kitaev anisotropy $J_z/J = 2.5$ and a hopping constant of $t/J=3.0$ is shown in Fig.~\ref{fig::arpesFM}. We observe that most parts of the spectral function resemble that of a free hole (green line), meaning that it follows the dispersion from only $H_t$ in Eq.~\eqref{eq::Kitaev_tJ}. This is consistent with a large FM cluster formation, where the hole can hop ballistically without distorting the spin background. Notably, both the energy dispersion and distribution of spectral weight $Z_\mathbf{k}$, obtained from this simple model, exhibit good agreement with the full spectrum of the Kitaev $t-J$ model. For the spectral weight, we fit a Gaussian to the low energy ($-\omega + \mu < 3t$) and high energy ($-\omega+\mu>3t$) branch of the MPS spectrum and integrate over the energies. The direct comparison of the spectral weight shows that indeed there are two branches in the spectrum, as for the free hopping case, but at each momentum, most of the weight is found in only one of them. Additionally, a more careful look at the spectral weight reveals missing weight in the sum of low and high energy branches for the MPS data, $Z_\mathbf{k}^{(-\omega+\mu>3t)}+Z_\mathbf{k}^{(-\omega+\mu<3t)} < 1$, even though the sum rule $\int d\omega A(\mathbf{k}, \omega)=1$ is fulfilled, when numerically integrating over the whole spectrum. This indicates that the spectrum also has some contributions from a continuum, which is especially visible in the $M_2$-$M_3$ cut of the spectrum. There, the broadening is much wider than the applied Gaussian broadening for the numerical algorithm. Hence, even though the free-hole picture accounts for most features in the spectrum, it does not fully describe all the details, where the continuum suggests further interactions between several individual constituents. Although a broad continuum is expected when the hole separates into fractional quasiparticles, these features here are too fragile to infer the spin-liquid nature of the ground state from the hole spectrum. By contrast, the hole spectrum directly uncovers the dynamically emerging ferromagnetism induced by dynamically doping the hole, which is an interesting phenomenon on its own.

\begin{figure}[t]
\centering
\includegraphics[trim={0cm 0cm 0cm 0cm},clip,width=0.95\linewidth]{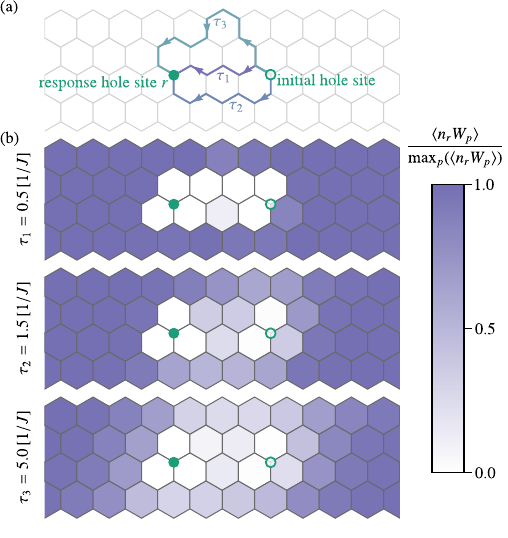}
\caption{\textbf{Flux dynamics after hole insertion.} (a) When hopping from the initial site to the response site $r$, the hole can move along different paths contributing to the dynamics at the corresponding times. (b)~The correlations $\langle n_r W_p \rangle$ between a hole at a fixed site $r$ and all plaquettes $p$ get destroyed along the paths as time evolves from $\tau_1$ to $\tau_3$. Shown data is for ferromagnetic Kitaev couplings $J>0$, $J_z/J = 2.5$ and $t/J=3.0$.}
\label{fig::fluxes}
\end{figure}

To further investigate the connection between the flux decay, as shown in Fig.~\ref{fig::nagaoka}~(c), and the hole motion, we look at the correlation functions $\langle n_r W_p \rangle$ for a fixed hole position $r$ six sites away from the original site of the hole insertion at time $\tau_0 = 0\,[1/J]$. 
Initially, the hole is positioned at a fixed site in the middle of the cylinder, resulting in an expectation value of zero for the hole everywhere else and hence leads to a vanishing correlator. As time progresses, the hole propagates towards site $r$ along various paths, as illustrated in Fig.~\ref{fig::fluxes}~(a). After a short time $\tau_1 \approx 0.5\,[1/J]$, the shortest path that is a direct connection from the initial sites reaches site $r$, leading to the reduction of $\langle W_p \rangle$ on plaquettes $p$ along that particular path. Subsequently, at longer times $\tau_2$ and $\tau_3$, additional paths become accessible, resulting in the decay of $\langle W_p \rangle$ on more plaquettes in a wider range around the sites involved. Since the flux $W_p$ decays in a large area between the hole and its initial position, we can disregard the possibility of a composite quasiparticle, such as bound flux-hole states, that could move around freely but restores the spin background after propagating further.

To summarize, for ferromagnetic Kitaev couplings $J>0$, the hole does not separate into fractionalized quasiparticles. Instead, the spin liquid state is destroyed, paving the way for a dynamically emerging ferromagnetic state. In the thermodynamic limit, only an infinitely fast hole is expected to give rise to an extensive FM polarization of the system. Our analysis, however, shows that already for intermediate hopping strength significant FM polarization clouds can be observed. Above, we focus on the specific gapped case of $J_z/J = 2.5$ and $t/J=3.0$. The same phenomena also occur for other ratios of $t/J$; data are shown in the ``Methods'' section. \\

\noindent \textbf{Antiferromagnetic Kitaev couplings: Fast~holes}

\begin{figure*}[t]
\centering
\includegraphics[trim={0cm 0cm 0cm 0cm},clip,width=1.0\linewidth]{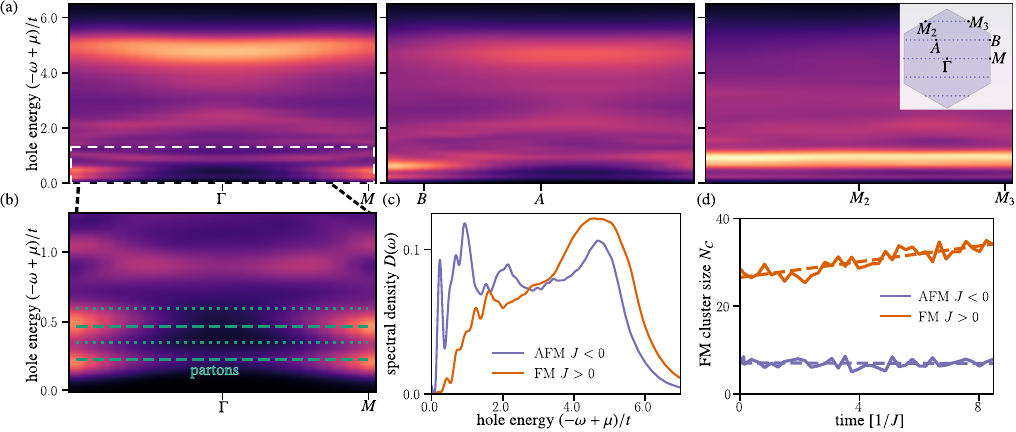}
\caption{\textbf{Hole spectral function for antiferromagnetic Kitaev couplings.} (a) The spectral function $A(\mathbf{k},\omega)$ for antiferromagnetic couplings, $J<0$, $J_z/J = 2.5$ and $t/|J|=3.0$ along the three distinct cuts in the Brillouin zone (see inset on the top right). (b) Zoom-in on the low-energy part of the spectrum shows several flat dispersions, which agree with the parton mean-field description (green dashed and dotted lines). (c) The spectral density $D(\omega) = \frac{1}{L_xL_y}\sum_\mathbf{k} A(\mathbf{k}, \omega)$, for antiferromagnetic (AFM $J<0$) and ferromagnetic (FM $J>0$) Kitaev couplings. (d) The FM cluster size $N_c$ increases for FM couplings but stays constant for AFM couplings, as further illustrated by the fits (dashed lines).}
\label{fig::arpesAFM}
\end{figure*}

\noindent Antiferromagnetic (AFM) Kitaev couplings $J<0$, which are predicted to be relevant for a recently explored class of Kitaev materials~\cite{Motome2020}, exhibit distinct behavior upon doping compared to ferromagnetic (FM) couplings, as demonstrated at the mean-field level~\cite{Okamoto2013, Scherer2014}. Thus, we also investigate the dynamics of a hole inserted into a Kitaev spin-liquid ground state with antiferromagnetic couplings. For very large values $t/|J|\rightarrow \infty$, Nagaoka's theorem is also applicable in this case, and an FM state will arise. However, as for the square lattice Heisenberg AFM, the hopping constant $t$ required for that may be very large because the gain in the kinetic energy of the hole has to be balanced with AFM spin fluctuations. Here, we focus on the strong coupling regime $t > |J|$ but avoid the Nagaoka FM for all considered values of the hopping strength.

In contrast to the FM case, the spectrum for AFM couplings exhibits a strikingly different character, Fig.~\ref{fig::arpesAFM}~(a). There is no clear signature of free hole hopping. Instead, the presence of several flat bands suggests the importance of Kitaev physics and the localization of excitations. At low energies $(-\omega+\mu)<1.5t$, the spectrum reveals a rich structure of dispersionless as well as dispersive bands; Fig.~\ref{fig::arpesAFM}~(b).

Analytically understanding the dynamics of AFM couplings proves to be challenging due to the intricate interplay between the hole, spin, and flux physics.
A possible approach to handle these different degrees of freedom is by fractionalizing the hole into a spinon and holon. Spinons couple to the spin degrees of freedom and hence contribute to the low-energy physics, while holons, carrying the charge quantum numbers, exhibit faster motion and contribute to high-energy features. The spectral function arises from the convolution of these two contributions~\cite{Senthil2008, Podolsky2009}:
\begin{equation}\label{eq::arpes_convolution}
    A(\kvec, \omega) = \int \mathrm{d}\nu \mathrm{d}\qvec \, A_{\mathrm{sp}}(\kvec-\qvec, \omega-\nu) A_{\mathrm{h}}(\qvec, \nu).
\end{equation}
On a one-dimensional lattice, the factorization of the spectral functions comes intrinsically from the spin-charge separation by mapping the $t$-$J$ Hamiltonian to independent spinon and holon parts in squeezed space~\cite{Bohrdt2018}. By contrast, in two dimensions, the decoupling of spinons and holons from the hopping term generally can only be applied on a mean-field level.
This approach yielded promising results for the chiral spin liquid~\cite{Kadow2022}. There, the lowest edge of the spectrum directly agrees with the dispersion of the corresponding spinon mean-field theory. Applying the same ansatz to the Kitaev spin liquid, we first rewrite the hole-doped Hamiltonian Eq.~\eqref{eq::Kitaev_tJ} in terms of holons and spinons, which interact with each other. Then, we perform a mean-field decoupling to extract the separate spectra of these fractionalized excitations; see ``Methods'' section for more details.

The convolution of spinons and holons features several flat bands similar to the MPS spectrum. These can be identified from the flat spinon bands corresponding to bond excitations of the $\chi^{x,y,z}$ Majorana fermions. For the anisotropic case $J_z/J=2.5$, the $z$-bands are at higher energy than the $x$- and $y$-bands. The holon contributes to the convolution with a renormalized free hopping, which has the most spectral weight at the two van Hove singularities. Therefore, the most prominent signatures in the combined spectrum are flat bands at energies of the spinon bands plus holons at the van Hove singularities. As a reference, we also put these energy lines into the low energy spectrum obtained from MPS; see dashed and dotted lines in Fig.~\ref{fig::arpesAFM}~(b) for the $x$-/$y$-spinons and $z$-spinon, respectively. 

\begin{figure*}
\centering
\includegraphics[trim={0cm 0cm 0cm 0cm},clip,width=1.0\linewidth]{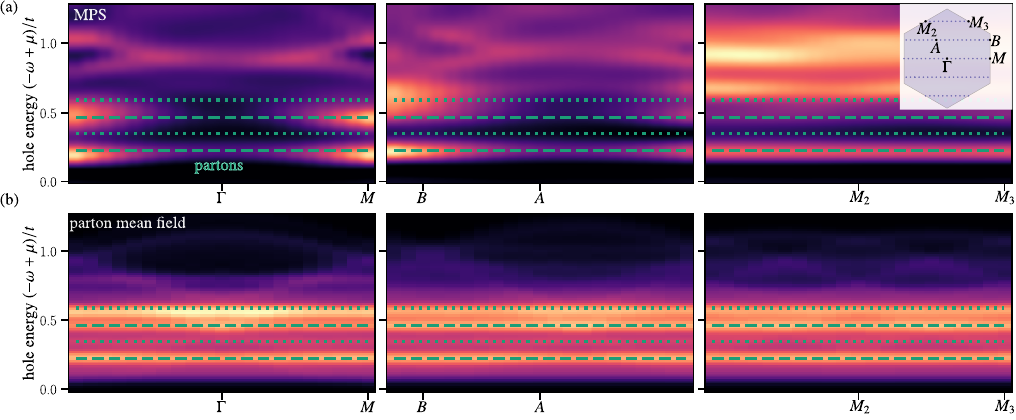}
\caption{\textbf{Hole spectral functions from MPS and parton mean-field theory.} Spectral function $A(\mathbf{k},\omega)$ at low energies for antiferromagnetic $J<0$, $J_z/J = 2.5$ and $t/|J|=3.0$ (a) from MPS time evolution and (b) from parton convolution; see Eq.~\eqref{eq::arpes_parton}. Dashed/ dotted lines correspond to energies of spinons for bond excitations along x- and y-bonds/ z-bonds, respectively, added to the van Hove points of the holons, which are expected to dominate the spectral response.}
\label{fig::parton_spectrum}
\end{figure*}

Overall, the parton mean-field theory captures the correct energies at which responses in the low-energy part of the hole spectrum are expected but fails to predict the spectral weight distribution; see Fig.~\ref{fig::parton_spectrum} in the ``Methods'' section. Therefore, the mean-field ansatz does not encompass all the relevant correlations and interplay between the charge and spin degrees of freedom. This shortcoming may be associated with the strong holon renormalization leading to effective hoppings $t_\mathrm{eff}^z\approx 0.18\,t$ along the $z$-bonds and $t_\mathrm{eff}^{x,y}\approx 0.11\,t$ along the $x$- and $y$-bonds; see ``Methods'' section for more details. Hence, the energy scales of spinons and holons are mixed but may be separated for even more extreme ratios $t/|J|$, which we cannot reliably access numerically.

To further compare the AFM couplings with FM couplings, we define the spectral density by integrating the spectrum over momentum $D(\omega) = \frac{1}{L_xL_y}\sum_\mathbf{k} A(\mathbf{k}, \omega)$; see Fig.~\ref{fig::arpesAFM}~(c). Each of the spectra is cut off at the edge of the free hole motion at $6\,t$ up to Gaussian broadening. However, while for FM couplings, the free hole dominates in the spectrum directly as a clear branch, for AFM couplings there are also flat bands in the high-energy regime of the spectrum, suggesting interactions of the free hole part with the flat band spinons, similarly to the renormalized holon at low energies.
Another striking difference is that for AFM couplings, a part of the weight is shifted towards lower energies, where there is almost no spectral weight in the FM case. The absence of weight for FM couplings is consistent with the free-hopping picture. Nevertheless, the total spectral density has some deviations from the one expected from only the free hole. Instead of expected sharp peaks at the van Hove points, $-\omega + \mu =2t$ and $-\omega + \mu =4t$, the signal is much more broadened, possibly due to further interactions between the hole and other constituents. The low-energy features present for the AFM couplings indicate that here the free hole does not describe the spectrum. To further support the absence of a FM polarization cloud around the hole, we conduct the same snapshot analysis as for the FM couplings. As shown in Fig.~\ref{fig::arpesAFM}~(d), the cluster size increases over time in the case of FM couplings. However, the cluster size remains constant at a low value for AFM couplings.

In the ``Methods'' section we present data for different hopping amplitudes $t/|J|$. The spectra look very similar for different hoppings and only differ slightly in the distribution of spectral weight. Similar to FM couplings, also for AFM couplings the overall energy scaling of all detected signatures is proportional to $t$ rather than $J$. This suggests a hole-dependent nature of these features and not only spin physics. By contrast, for the chiral spin liquid on the triangular lattice, the main energy scaling was found to be $\propto J$~\cite{Kadow2022}.
These observations let us conclude that for AFM couplings, holes, spins, and fluxes mutually influence each other, leading to the intricate dynamical interplay between the (fractionalized) excitations. The main reason for that is the strong renormalization of the hopping strength $t_\text{eff}$, which is no longer larger than the spin energy scales. Nevertheless, the hole spectrum shows signatures of flat spinon bands that are a direct consequence of the underlying Kitaev spin liquid and can therefore be used as a characteristic of this phase in ARPES experiments.\\

\noindent \textbf{Antiferromagnetic Kitaev couplings: Slow-hole~limit}

\begin{figure}[t]
\centering
\includegraphics[trim={0cm 0cm 0cm 0cm},clip,width=0.95\linewidth]{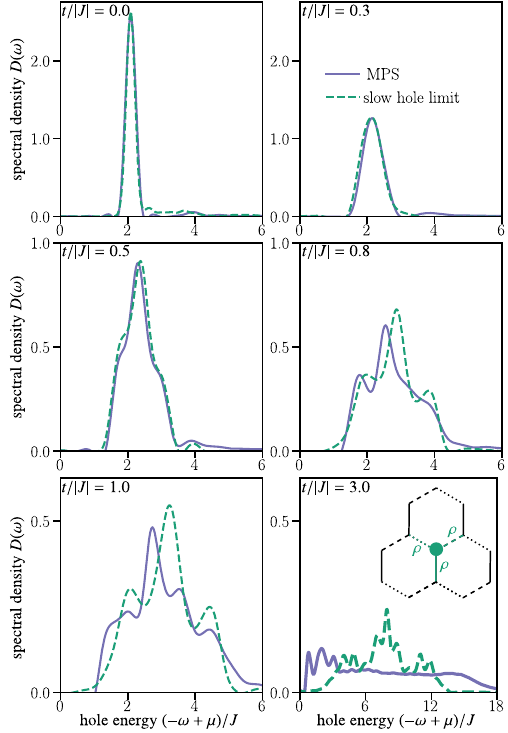}
\caption{\textbf{Slow-hole approximation.} The spectral density $D(\omega)=\frac{1}{L_xL_y}\sum_\mathbf{k} A(\mathbf{k}, \omega)$ obtained from MPS time evolution (purple) is compared with the slow-hole approximation (green) as described in the main text and sketched in the inset of the last panel. We fix antiferromagnetic couplings $J<0$, $J_z/J=2.5$ and vary $t/|J|$ from $0.0$ (exact stationary hole limit) to $3.0$ (fast holes). For $t/|J|=3.0$, a different energy scale of the x-axis is used in order to show the whole spectrum.}
\label{fig::slow_holes}
\end{figure}

\noindent Previous works have focused on the slow-hole limit $t\ll |J|$ as well~\cite{Halasz2014, Halasz2016}. For a stationary hole $t=0$, one can think of it as a vacancy site, where holes are described within the spin Hamiltonian Eq.~\eqref{eq::Kitaev}, i.e., the spin is not removed from the ``hole site'', but all interactions with neighboring spins are just turned off. We now want to investigate this limit to understand the limitations and merits compared to our numerical simulations.

We start with the description of a stationary hole $t=0$ and solve the quadratic Hamiltonian Eq.~\eqref{eq::Kitaev_majorana} by doubling the number of Majorana fermions to obtain complex fermions: 
\begin{equation}
    f_{i\in A} = \frac{1}{2} (\chi^0_i + i \underline{\chi}^0_i), \quad f_{i\in B} = \frac{i}{2} (\chi^0_i + i \underline{\chi}^0_i),
\end{equation}
where $\underline{\chi}^0_i$ are copies of the original Majorana fermions~\cite{Halasz2016}. From this, we obtain the doubled Hamiltonian,
\begin{align}\label{eq::Kitaev_doubled}
    H_d = H_K + \underline{H}_K &= i\sum_{a, \langle i j\rangle_a} J_a \hat u_{\langle i j \rangle_a} \left(\chi^0_i \chi^0_j + \underline{\chi}^0_i \underline{\chi}^0_j \right), \\
    &= \sum_{a, \langle i j\rangle_a} 2 J_a \hat u_{\langle i j \rangle_a} \left(\fd{i}\fo{j} + \mathrm{h.c.} \right).
\end{align}
We introduce a hole at site $j_0$ by switching off all interactions at the corresponding bonds,
\begin{equation}\label{eq::Kitaev_doubled_hole}
    \tilde{H}_d^{j_0} = H_d - \sum_{a, j \in \langle j_0 j\rangle_a} 2 J_a \hat u_{\langle j_0 j \rangle_a} \left(\fd{j_0}\fo{j} + \mathrm{h.c.} \right).
\end{equation}
Let $\ket{\Omega}$ and $\ket{\tilde{\Omega}}$ denote the ground states of the pure Kitaev model and the one with a stationary hole, respectively. We can compute the corresponding quasiparticle weight of the hole spectrum $Z=|\braket{\tilde{\Omega}|\Omega}|^2$ directly on the same cylinder geometry that we use for the MPS time evolution. To obtain the full spectrum for a stationary hole, we expand the hole Hamiltonian in the eigenbasis at half-filling, $\tilde{H}_d^{j_0} = \sum_i \tilde{\varepsilon}_i \ket{\tilde{\alpha}_i}\bra{\tilde{\alpha}_i}$,
\begin{equation}
    A(\omega) = \sum_i \delta(\omega-\tilde{\varepsilon}_i) |\braket{\tilde{\alpha}_i|\Omega}|^2.
\end{equation}
The spectral function is momentum independent since the hole is localized. In general, the sum above runs over exponentially many states. However, by evaluating the sum rule $\int d\omega A(\omega)=1$, we estimate that including only one-particle excitations on top of the ground state already account for most of the spectral weight~\cite{Knolle2014}, e.g., for $J_z/J=2.5$, we find $\int d\omega A_1(\omega)=0.997$. Therefore, when comparing this exact result to the MPS spectral function, we find excellent agreement up to little numerical deviations for the immobile hole; see Fig.~\ref{fig::slow_holes}. This serves as a stringent benchmark for our numerical approach. 

To include a finite but small hopping amplitude $t$ for the holes, we focus on the gapped phase $J_z/J>2$. In that limit, the effective hole hopping was derived in Ref.~\cite{Halasz2014}. This derivation is strictly valid as long as the hole excitations do not couple to any bulk excitations of the undoped Kitaev model, i.e., for $t/|J| \ll (J/J_z)^4$ (the flux gap). Then the holes are associated with flux, fermion, and plaquette quantum numbers. In our case of a single hole, the former two have to be trivial, while the plaquette quantum number $p=0,1$ implies a degeneracy for the ground state $\ket{\tilde{\Omega}}$. We find, however, that all the calculations presented below yield identical results for both values. In the anisotropic limit, the hopping amplitude gets renormalized along the different bonds as following $t_x=t/2$, $t_y=t/2$, and $t_z=t$~\cite{Halasz2014}. This can be intuitively understood from the isolated dimer limit $J_z/J\rightarrow \infty$, where the ground state consists of dimers on the $z$-bonds $(\ket{\uparrow\downarrow}_z+\ket{\downarrow\uparrow}_z)/\sqrt{2}$. Inserting either an up- or down-hole on one lattice site will break up the corresponding dimer, e.g., $\ket{\uparrow\circ}_z$. The hole can hop freely back and forth along the same dimer on the $z$-bond $\ket{\uparrow\circ}_z \rightarrow \ket{\circ\uparrow}_z$. Hopping along the $x$- or $y$-bonds requires hopping to a neighboring dimer
\[ \ket{\uparrow\circ}_z \otimes (\ket{\uparrow\downarrow}_{\tilde{z}}+\ket{\downarrow\uparrow}_{\tilde{z}})/\sqrt{2} \rightarrow (\ket{\uparrow\uparrow}_z \otimes \ket{\circ\downarrow}_{\tilde{z}} + \ket{\uparrow\downarrow}_z \otimes \ket{\circ\uparrow}_{\tilde{z}})/\sqrt{2}\] and therefore, effectively reduces the hopping amplitude by a factor of $1/2$, because of the strong ferromagnetic energy penalty $J_z/J\rightarrow \infty$ in the first term after hopping. Away from the isolated dimer limit, the renormalization changes little as long as $J_z/J>2$~\cite{Halasz2014}. 

The doped Hamiltonian Eq.~\eqref{eq::Kitaev_doubled_hole} is only exactly solvable for a specific hole configuration. To interpolate between the limit of a completely stationary hole and a free moving one in the translational invariant Kitaev ground state, a variational ansatz was put forward~\cite{Halasz2016}, $H(\rho) = \rho H_d + (1-\rho) \tilde{H}_d^{j_0}$; sketched in the inset of Fig.~\ref{fig::slow_holes}. Here, we modify this ansatz and allow for the bond strength to vary over all lattice sites $\rho \mapsto \mathbf{\rho}(\{j \})$. The interpretation follows straightforwardly from the stationary picture. As the hole moves through the system, at different times, different bonds will be modified depending on the probability of finding the hole at the corresponding site. According to the hole expectation value $\langle n_j \rangle$, we choose $\rho_j = 1 - \langle n_j \rangle$.

To describe a local hole that spreads dynamically, we approximate the time evolution by step-wise adjusting $\langle n_j(\tau) \rangle$ and, hence, also $\rho_j(\tau)$, where $\langle n_j(\tau) \rangle$ is determined by a free hopping hole with $t_x=t/2$, $t_y=t/2$, and $t_z=t$ according to the slow hole prediction. Therefore, we get a Trotterized time evolution for the response function with step-size $\delta=\tau/N$,
\begin{equation}
    C_{ij}(\tau) \approx e^{iE_0\tau} \bra{\Omega} a_i^{ } \e^{-iH_N\delta} \e^{-iH_{N-1}\delta} \dots \e^{-iH_2\delta} \e^{-iH_1\delta} a_j^\dagger\ket{\Omega},
\end{equation}
where we introduce hole operators $a_j^{(\dagger)}$ which locally switch off the couplings. Here, $E_0$ is the ground state energy of the undoped Kitaev model and for each $H_n$ ($n=1, \dots ,N$) we modify $\rho$ as described before according to the corresponding hole expectation at time $n\cdot\delta$. Since all $H_n$ are quadratic fermionic Hamiltonians, we can diagonalize them $H_n = \sum_i \varepsilon_i^{(n)} \ket{\alpha_i^{(n)}}\bra{\alpha_i^{(n)}}$. By insertion of identities after each time step, the correlation function simplifies to a product of overlaps,
\begin{align}
    C_{ij}(\tau) \approx e^{iE_0\tau} \sum_{i_1, \dots i_N} & \bra{\Omega} a_i^{ } \ket{\alpha_{i_N}^{(N)}} \e^{-i\delta\varepsilon_{i_N}^{(N)}} \braket{\alpha_{i_N}^{(N)}| \alpha_{i_{N-1}}^{(N-1)}}\nonumber\\
    &\dots \braket{\alpha_{i_2}^{(2)}|\alpha_{i_{1}}^{(1)}} e^{-i\delta\varepsilon_{i_1}^{(1)}} \bra{\alpha_{i_{1}}^{(1)}} a_j^\dagger\ket{\Omega}.
\end{align}
In practice, again we do not need to sum over all eigenstates $\ket{\alpha_{i_n}^{(n)}}$: Because the Hamiltonians change only slightly from one time step to the other, it is sufficient to take the overlap with only the ground states $\ket{\Omega^{(n)}}$. We find that indeed the overlaps are very close to one. Thus, we can simplify the time-dependent correlations to
\begin{equation}
    C_{ij}(\tau) \approx \bra{\Omega} a_i^{ } \ket{\Omega^{(N)}} \prod_{n=1}^{N-1} \left( e^{-i\delta\varepsilon_{0}^{(n)}} \braket{\Omega^{(n+1)} | \Omega^{(n)} } \right) \bra{\Omega^{(1)}} a_j^\dagger\ket{\Omega}.
\end{equation}
The spectral function $A(\mathbf{k}, \omega)$ is then computed by spatial and temporal Fourier transformations of the time-dependent correlation functions, similar to the evaluation of the data from the MPS time evolution.

We compare the slow-hole approximations to the spectra obtained from the full MPS time evolution in Fig.~\ref{fig::slow_holes}. For $t/|J|=0.0$, the analytic description of stationary holes is exact, and we find that the spectra indeed are almost identical; demonstrating that our numerical results are well converged as discussed above. When increasing the hopping strength, we see a remarkable resemblance between the slow-hole limit and the MPS data. Note that the formally required limit for the approximation to be valid is $t/|J| \ll (J/J_z)^4 \approx 0.025$. This ensures that the hole does not couple to the flux excitation of the bulk.
Yet, our numerics agree well up to $t/|J|\approx 0.5$. This indicates that even at higher energy, the hole-flux interactions are not relevant to the shape of the spectrum. Instead, the hole is best described by vacancy sites that spread slowly over the whole system and modify the Kitaev spin-liquid ground state only slightly.
Upon increasing the hopping strength $t/|J| \gtrsim 0.8$, the general shape between both curves is still similar, but small deviations between peak positions occur. Furthermore, for $t/|J| \gtrsim 1.0$ additional peaks in the MPS spectrum appear. This suggests that further modes in the model with significant spectral weight are excited. Since this can include complex interactions between the hole and flux or matter excitations in the Kitaev model, the simple ansatz for the slow holes cannot capture these features anymore. Eventually, at even faster hopping $t/|J| \gtrsim 3.0$, the spectra look very different, meaning that the slow-hole picture breaks down, as expected. We find that the overall bandwidth is not given by the renormalized hoppings $t/2$ as predicted from the slow hole ansatz but by the full bandwidth $\sim t$. Moreover, we see a shift of spectral weight towards low energies, where distinct peaks are visible, which transition into a broad continuum at high energies, and the parton ansatz becomes more reasonable.\\

\noindent \textbf{Local scanning tunneling microscopy spectra}

\begin{figure}[t]
\centering
\includegraphics[trim={0cm 0cm 0cm 0cm},clip,width=0.95\linewidth]{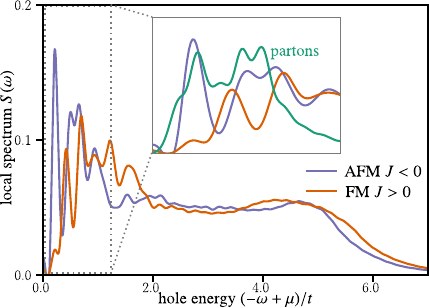}
\caption{\textbf{Local spectra.} The local spectral function $S(\omega)$, Eq. \eqref{eq::local_spectrum}, as measured by scanning tunneling microscopy (STM), is shown for ferromagnetic couplings $J>0$ and antiferromagnetic couplings $J<0$; both with $J_z/J = 2.5$ and $t/|J|=3.0$. The inset focuses on the low energy regime that is compared to the parton ansatz (green line).}
\label{fig::local_spectrum}
\end{figure}

\noindent So far, we focused on the momentum-resolved and energy-resolved hole spectrum as measured by ARPES experiments, Eq.~\eqref{eq::arpes} or their sum rules. Here, we will consider the spatially averaged local hole spectrum, that is measurable with scanning tunneling microscopy (STM):
\begin{equation}
    S(\omega) = \frac{1}{L_xL_y} \sum_{\sigma=\uparrow,\downarrow} \sum_j \int \mathrm{d} \tau \e^{i\tau \omega} \bra{\psi_0}\cd{j\sigma}(\tau) \co{j\sigma}(0) \ket{\psi_0}.
\label{eq::local_spectrum}
\end{equation}
Although for lattices with a single site per unit cell $S(\omega)$ is equal to the spectral density $D(\omega)= \frac{1}{L_xL_y}\sum_\mathbf{k} A(\mathbf{k}, \omega)$, this is not the case for the honeycomb lattice, which has a two-sublattice structure; see methods for details. 
Recent experimental proposals suggest that tunneling through a Kitaev spin liquid layer into a metal beneath will lead to a distinct response of the underlying spin liquid~\cite{Feldmeier2020, Koenig2020, Kao2023}. Here, we consider a different scenario, where tunneling occurs directly in and out of the spin liquid layer by injecting mobile holes.

The local spectra for both FM and AFM couplings stretch over a broad continuum bounded by the free-hole bandwidth of $6\,t$; see Fig.~\ref{fig::local_spectrum}. In contrast to the ARPES results, the local spectra $S(\omega)$ of the FM possesses additional structure at low energies. These additional peaks cannot be explained with the free hole alone but indicate the additional dressing by local dynamical spin correlations.
For AFM couplings, the low-energy peak agrees reasonably with the parton mean-field ansatz derived in the ``Methods'' section; see the green line in the inset of Fig.~\ref{fig::local_spectrum}. This supports the concept of fractionalized quasiparticles in this energy region of the spectrum and offers a different probe to detect signatures of the Kitaev spin liquid.\\

\noindent \textbf{\large Discussion}\\
\noindent Our results reveal an intriguing behavior of a mobile hole doped into the Kitaev spin-liquid ground state. We looked at the real-time dynamics of the hole spreading and studied the energy and momentum resolved hole spectra, as measured by ARPES as well as the local spectral function, as obtained from STM experiments. Our simulations were carried out on finite cylinder geometries using tensor network methods, however, we expect at least qualitatively similar behavior for the 2D limit.

The ARPES response is drastically different for ferromagnetic (FM) and antiferromagnetic (AFM) Kitaev couplings between neighboring spins.
The FM case is characterized by the formation of clusters of aligned spins leading to a dynamical Nagaoka ferromagnet that allows for coherent hole hopping. Intriguingly, the STM spectra show coherent dynamical features even at low energies, which are not present in ARPES. Conversely, AFM couplings robustly show multiple dispersionless features at low energies. In the slow-hole limit, we find that the hole does not couple to any other excitations and is described as a vacancy that modifies the spin liquid. However, when increasing the hopping strength, flux and matter excitations of the Kitaev model become relevant and significantly alter the spectral function. Understanding these modifications is challenging since the exact solution of the Kitaev model is lost and the hole interacts non-trivially with the spin background. A parton mean-field ansatz gives some insights into the spectral responses for AFM couplings but fails to reproduce the proper spectral weight along the different momentum cuts in the Brillouin zone. Thus, interactions beyond a parton mean-field theory are required in the considered parameter regime and may be investigated in more detail in future studies.

Our work shows that doping a spin-liquid ground state with a single hole can have several different outcomes. In one scenario, the hole can fractionalize into a slow spinon and a fast holon. In that case, the low-energy spectrum directly probes the spinon dynamics of spin liquids~\cite{Senthil2008, Podolsky2009, Kadow2022}. In the other scenario, the hole significantly modifies the spin-liquid ground state in its vicinity and dynamically forms a Nagaoka ferromagnet around the hole. Both scenarios are realized in the doped Kitaev model depending on the sign of spin interactions.
This behavior should be contrasted to the spectral functions of Mott insulating states with magnetic order as, for example, described by the $t-J$ model on the square lattice. On the one hand, the AFM hole spectrum exhibits a quasi-particle peak at low energies that can be effectively described by a bound state of a holon and a spinon~\cite{Bohrdt2020}. In contrast, for the Kitaev spin liquid, we assume the holon and spinon to be deconfined. On the other hand, a FM spin exchange $J$ results in a FM ground state on the square lattice. The single-hole problem can be solved exactly since hole hopping leaves the spin background unmodified and describes the hopping of a free hole. This is in sharp contrast to our results for the Kitaev spin liquid, where the ground state does not have any order initially. Instead, the FM forms dynamically and, along with it, the effective free-hole-like response in the spectral function.

It will be interesting to see in which category the response of other slightly doped spin liquids may fall.
This also raises the question of the general stability of quantum spin liquids upon hole doping. Tensor network methods for ground states as well as dynamical response functions offer one possible route to study the structure of the phases that could be realized in the doped Kitaev model~\cite{Hui-Ke2023, Peng2021}. Furthermore, they may shed light on which perturbations can stabilize topological superconductivity beyond the mean-field approach.\\

\noindent \textbf{\large Methods}

\noindent \textbf{Physical spins and form factor}

\noindent In the promising candidate materials exhibiting strong Kitaev interactions, such as 5$d$ iridate compounds Na$_2$IrO$_3$ and 4$d$ $\alpha$-RuCl$_3$, the partially filled d$^5$ orbital is split into $e_g$ and $t_{2g}$ ($|xy\rangle$, $|yz\rangle$, and $|zx\rangle$) orbitals. Because of the strong spin-orbit coupling, the $t_{2g}$ multiplet is further divided into a $J_\mathrm{eff}=3/2$ quartet and a $J_\mathrm{eff}=1/2$ Kramers doublet. The low-energy physics is dominated by the $J_\mathrm{eff}=1/2$ doublet with a reduced bandwidth. Here, we would like to demonstrate the relation between hybridized pseudo-spins and physical spins in realistic materials such as Na$_2$IrO$_{3}$.

The Kramers doublet, $(c^\dag_{\uparrow}, c^\dag_{\downarrow})$, can be expressed in terms of $t_{2g}$ orbitals as~\cite{Shitade2009}
\begin{equation}
\begin{split}
&c^\dag_{\uparrow}=\frac{1}{\sqrt{3}}\left(d^\dag_{xy,\uparrow}+d^\dag_{yz,\downarrow}+id^\dag_{zx,\downarrow}\right)\\
&c^\dag_{\downarrow}=\frac{1}{\sqrt{3}}\left(-d^\dag_{xy,\downarrow}+d^\dag_{yz,\uparrow}-id^\dag_{zx,\uparrow}\right),
\end{split}
\end{equation}
where $d^{\dag}_{m,\sigma}$ is an electron for orbital $m\in(xy,yz,zx)$ and spin $\sigma\in(\uparrow,\downarrow)$ and the lattice site index has been omitted. We can introduce a vector of electrons 
${\bf d}^\dag_{}=\left(d^\dag_{xy,\uparrow},d^\dag_{xy,\downarrow},d^\dag_{yz,\uparrow},d^\dag_{yz,\downarrow},d^\dag_{zx,\uparrow},d^\dag_{zx,\downarrow}\right)$ and define a projector $P$
\begin{equation}
    P=\frac{1}{\sqrt{3}}\left(\begin{array}{cccccc}
    1 & 0 & 0 & 1 & i\\
    0 & -1 & 1 & 0 & -i & 0 
    \end{array}\right)^T.
\end{equation}
Then we obtain a compact form of $(c^\dag_{\uparrow}, c^\dag_{\downarrow})={\bf d}^\dag_{}P$.

The physical spin operators at site $j$ are defined as
\begin{equation}
    S^a_j = \sum_{m=(xy,yz,zx)}\sum_{s,s'}d^\dag_{j,m,s}\sigma^a_{ss'}d^\dag_{j,m,s'}.
\end{equation}
Now we would like to project the spin to the low-energy $J_\text{eff}=1/2$ bands. In order to do this, we first need to find the null space corresponding to the projector $P$, dubbed $P_{\perp}$. Note that $P_{\perp}$, whose explicit form is not important, can be obtained by extracting the eigenstates of zero modes of $PP^\dag$.
Indeed, those zero modes of $PP^\dag$ correspond to the basis of the high-energy $J_\text{eff}=3/2$ quartet, and we denote those modes as $(c^\dag_{1},...,c^\dag_{4})$. Using $P$ and $P_{\perp}$, one can express the electrons as ${\bf d}^\dag={\bf c}^\dag(P+P_{\perp})^\dag$ where ${\bf c}^\dag\equiv 
(c^\dag_{\uparrow},c^\dag_{\downarrow},c^\dag_{1},...,c^\dag_{4})$. Therefore, we can rewrite the physical spin operators in terms of $c$-fermions, and after projecting into the low-energy doublet sector, we find that
\begin{equation}
    S^a_j\rightarrow{}-\frac{1}{3}\sum_{s,s'}c^\dag_{s}\sigma^a_{ss'}c^{}_{s'}=-\frac{1}{3}\sigma^a_j. 
\end{equation}

\begin{figure}[t]
\centering
\includegraphics[trim={0cm 0cm 0cm 0cm},clip,width=0.95\linewidth]{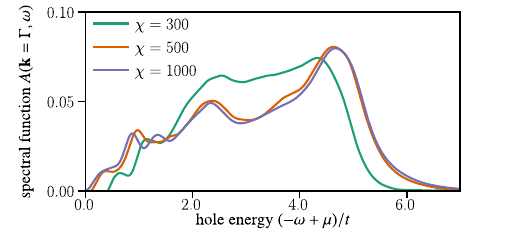}
\caption{\textbf{Convergence of the spectral function.} The spectral function $A(\mathbf{k}, \omega)$ for ferromagnetic couplings, $J>0$, $J_z/J = 2.5$ and $t/J=3.0$ at the $\Gamma$ point is converged for larger values of the maximal MPS bond dimension $\chi>500$, but still shows considerable deviations for $\chi=300$.}
\label{fig::bond_dimension}
\end{figure}

The tensor of the dynamical hole spectral function is thus defined as 
\begin{equation}
    \mathcal{A}_{ms,m's'}({\bf k},\omega)=\int{}{\rm d} \tau{}e^{i\tau\omega}\langle{}\psi_0|d^\dag_{{\bf k}, ms}(\tau)d^{}_{{\bf k}, m's'}(0)|\psi_0\rangle.
\end{equation}
By viewing $\mathcal{A}({\bf k},\omega)$ as a $6\times6$ matrix, it is related to the spectral function for $c$-fermions by a unitary transformation as $\mathcal{A}({\bf k},\omega)=(P+P_{\perp})^*\mathcal{F}({\bf k},\omega)(P+P_{\perp})^T$, where,
\begin{equation}
\begin{split}
    \mathcal{F}_{\mu,\nu}({\bf k},\omega)&=\int{}{\rm d} \tau{}e^{i\tau\omega}\langle{}\psi_0|c^\dag_{{\bf k}, \mu}(\tau)c^{}_{{\bf k}, \nu}(0)|\psi_0\rangle
\end{split}
\end{equation}
with $H$ the Hamiltonian and $E_0$ the ground-state energy. We mainly focus on the excitations with energy being within $E_{W}$, the bandwidth of the $J_\text{eff}=1/2$ band. Within the frequency regime of $\omega < E_{W}$ ($\hbar = 1$), we approximately have 
\begin{equation}
   \mathcal{F} \approx \left( \begin{array}{ccc}
        F_{\uparrow,\uparrow} & F_{\uparrow,\downarrow} & 0\\
        F_{\downarrow,\uparrow} & F_{\downarrow,\downarrow} & 0\\
        0 & 0 & 0_{4\times4}
    \end{array}\right).
\end{equation}
Consequently, the orbital and spin resolved hole spectral function can be obtained by only computing the $2\times{}2$ spectral function for the pseudo spins $F_{\sigma,\sigma'}$, as
\begin{equation}
    \mathcal{A}=\frac{1}{3}\left(
        \begin{array}{cccccc}
         F_{\uparrow,\uparrow} &	-F_{\uparrow,\downarrow} & F_{\uparrow,\downarrow}	& F_{\uparrow,\uparrow} & -iF_{\uparrow,\downarrow} & iF_{\uparrow,\uparrow} \\
     -F_{\downarrow,\uparrow} &	F_{\downarrow,\downarrow} & -F_{\downarrow,\downarrow}	& -F_{\downarrow,\uparrow} & iF_{\downarrow,\downarrow} & -iF_{\downarrow,\uparrow} \\
     F_{\downarrow,\uparrow} & -F_{\downarrow,\downarrow} & F_{\downarrow,\downarrow} & F_{\downarrow,\uparrow} & -iF_{\downarrow,\downarrow} & iF_{\downarrow,\uparrow}\\
 F_{\uparrow,\uparrow} & -F_{\uparrow,\downarrow} & F_{\uparrow,\downarrow} & F_{\uparrow,\uparrow} & -i F_{\uparrow,\downarrow} & i F_{\uparrow,\uparrow} \\
iF_{\downarrow,\uparrow} & -i F_{\downarrow,\downarrow} & i F_{\downarrow,\downarrow} & i F_{\downarrow,\uparrow} & F_{\downarrow,\downarrow} & -F_{\downarrow,\uparrow} \\
 -i F_{\uparrow,\uparrow} & i F_{\uparrow,\downarrow} & -i F_{\uparrow,\downarrow} & -i F_{\uparrow,\uparrow} & -F_{\uparrow,\downarrow} & F_{\uparrow,\uparrow} \\
     \end{array}\right).
\end{equation}
In the end, the energy- and momentum-resolved spectral function defined in the main text Eq.~\eqref{eq::arpes} is just the trace of the above matrix, as
\begin{equation}
    A({\bf k},\omega)=F_{\uparrow,\uparrow}({\bf k},\omega) + F_{\downarrow,\downarrow}({\bf k},\omega).
\end{equation}

\begin{figure*}[t]
\centering
\includegraphics[trim={0cm 0cm 0cm 0cm},clip,width=1.0\linewidth]{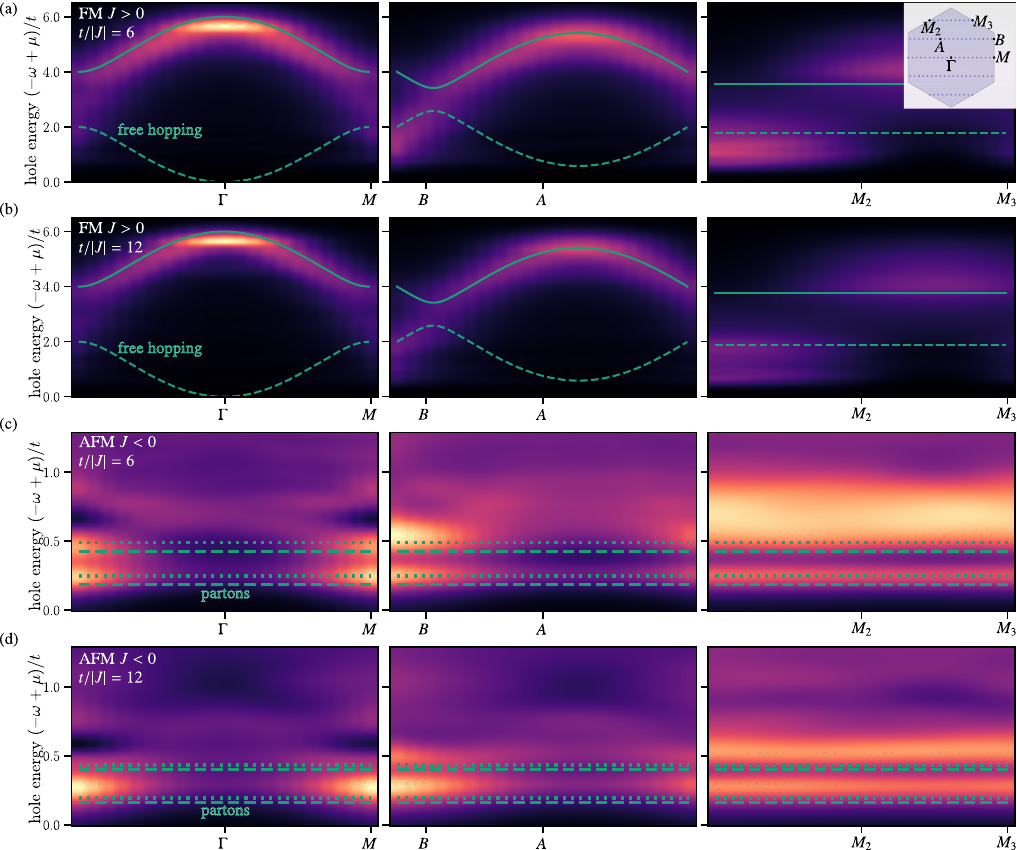}
\caption{\textbf{Additional data for hole spectral functions with hopping amplitudes $t/|J|=6,$ and $12$.} (a),~(b) For ferromagnetic Kitaev couplings ($J>0$), the spectrum mainly consists of the response of a free hole (green lines). (c),~(d) Antiferromagnetic Kitaev couplings ($J<0$) show only slight changes at low energies for different values for $t/|J|$. Spectra are compared to the parton ansatz; dashed and dotted lines for spinon excitations along x- or y-bonds and z-bond, respectively. Note that all energy scales are in units of $t$, and we fix $J_z/J=2.5$.}
\label{fig::spectra}
\end{figure*}

\begin{figure*}[t]
\centering
\includegraphics[trim={0cm 0cm 0cm 0cm},clip,width=1.0\linewidth]{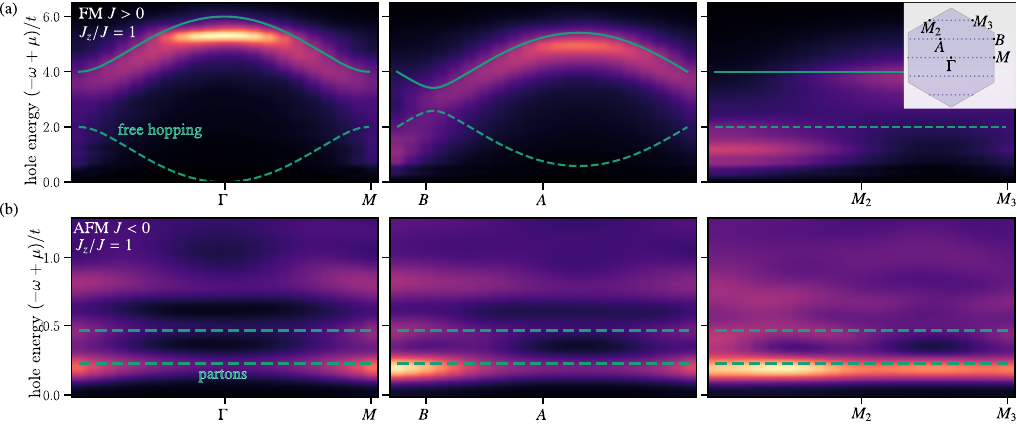}
\caption{\textbf{Additional data for hole spectral functions with isotropic Kitaev couplings $J_z/J=1$.} (a) For ferromagnetic Kitaev couplings ($J>0$), the main contribution to the spectrum resembles the response of a free hole (green lines). (b) For antiferromagnetic Kitaev couplings ($J<0$), the first flat features can be described by the parton ansatz (dashed green lines). We fix $t/|J|=3.0$.}
\label{fig::spectra_iso}
\end{figure*}

Hence, the only contributions come from terms that do not mix the different spins. Numerically, we find that $F_{\uparrow,\uparrow}({\bf k},\omega) = F_{\downarrow,\downarrow}({\bf k},\omega)$ for all investigated parameters. Therefore, spin-resolved spectra provide only additional information when the material breaks spin-symmetry. \\

\noindent \textbf{Details on the numerical methods}

\noindent We use MPS methods for computing the time evolution of the state after hole injection. All simulations were performed with the Python library TeNPy \cite{Hauschild2018}. Numerical costs grow linearly in $L_x$ but exponentially in $L_y$. Thus, we restrict ourselves to cylinders with $L_y=4$ unit cells (i.e., eight sites) and $L_x=20$. To reach the large bond dimensions of up to $\chi=1000$ required to obtain converged results, we have utilized $\mathrm{U}(1)\times \mathbb{Z}_2$ symmetries for particle number and spin parity conservation, respectively.

To study the hole dynamics, we first find the approximate ground state of the pure spin model Eq.~\eqref{eq::Kitaev} with DMRG. Several flux sectors are degenerate on the chosen cylindrical geometry with open boundary conditions along the $x$-direction. We favor the flux-free sector explicitly as a ground state in our simulation by adding plaquette terms $\propto -\sum_p W_p$. For a periodic or infinite system, it is known that the ground state has to be flux-free~\cite{Lieb1994}.

The time evolution was realized via an MPO representation of the time evolution operator, the $W_{\mathrm{II}}$ operator from Ref.~\cite{Zaletel2014}. To apply the operator to the state, we have first to use the more costly zip-up methods for a few time steps to avoid being stuck in a local minima before continuing with a variational truncation scheme with fixed MPS bond dimension $\chi$. Depending on the hole hopping we choose step size $\delta\tau = 0.05/\,t$ for fast holes $t/|J|>1.0$ and $\delta\tau = 0.025/\,|J|$ otherwise.

In this way, we compute a time evolution after injecting a hole into the ground state $\ket{\psi_0}$ of the pure Kitaev model
\begin{equation}
    \ket{\psi(\tau)} = \e^{-iH\tau} \co{j\sigma}\ket{\psi_0},
\end{equation}
where $j$ is a site in the middle of the cylinder. We check that the cylinder is long enough ($L_x=20$) such that excitations do not reach the boundaries on the simulated time scales to limit the boundary effects.

To access the spectral function, we employ established MPS methods~\cite{Paeckel2019}. Initially, we obtain an MPS approximation for the ground state without hole $\ket{\psi_0}$ through the DMRG algorithm. The calculation is performed on a finite $L_y\times L_x$ system, with $L_y=4$ and $L_x=20$. For that state, we compute the time-dependent correlation function 
\begin{equation}\label{eq::Cij}
    C_{ij}(\tau) = \sum_\sigma \braket{\psi_0| \e^{i \tau H} \cd{j\sigma} \e^{-i \tau H} \co{i\sigma}|\psi_0}, \quad i,j \in \{A,B\}.
\end{equation}
Using translational invariance of the ground state, we only need to compute two time evolutions after hole insertion at one site in each of the sublattices $A$ and $B$ of the honeycomb lattice. For ARPES, the hole can be ejected by the photon on any of the sublattices. Therefore, the measured spectral weight is only periodic in an extended Brillouin zone~\cite{Luscher2006, Trousselet2014} and the corresponding Fourier transformation of the annihilation operator is given by 
\begin{equation}\label{eq::FourierARPES}
    \co{\kvec\sigma} = \frac{1}{\sqrt{N}} \sum_{i \in \{A,B\}} \e^{i\kvec \rvec_i} \co{i\sigma}.
\end{equation}
Here, the sum includes all lattice sites. This has to be contrasted to the usual definition of the sublattice Fourier transformation, which is for instance used to compute the dispersion relations for the parton mean-field ansatz below
\begin{equation}\label{eq::FourierSub}
    \co{\kvec,A(B) \sigma} = \frac{1}{\sqrt{N}} \sum_{i \in A (B)} \e^{i\kvec \rvec_i} \co{i,A (B)\sigma}.
\end{equation}
For Eq.~\eqref{eq::FourierSub}, a unique inverse Fourier transformation exists. In contrast, for the operator in Eq.~\eqref{eq::FourierARPES}, we have to distinguish between $\kvec \in \mathrm{1. BZ}$ and $\kvec \notin \mathrm{1. BZ}$. These two cases are related to the sublattice Fourier transformation by:
\begin{equation}\label{eq::FourierBZ}
    \co{\kvec \sigma} = 
    \begin{cases}
        \co{\kvec,A \sigma} + \co{\kvec,B \sigma} & \mathrm{for\,} \kvec \in \mathrm{1. BZ} \\
        \co{\kvec,A \sigma} + \e^{i\mathbf{g}\rvec_{AB}}\co{\kvec,B \sigma} & \mathrm{for\,} \kvec-\mathbf{g} \in \mathrm{1. BZ}
    \end{cases},
\end{equation}
where $\mathbf{g}$ is the reciprocal lattice vector that brings $\kvec$ back to the first Brillouin zone and $\rvec_{AB}$ is the vector connecting the two sites in the unit cell. We use these conventions to compute the spatial Fourier transformation of Eq.~\eqref{eq::Cij}.
The subtlety of contributions from both sublattices also leads to the difference between the local hole spectrum and the integrated spectral density $S(\omega)\neq D(\omega)$, see Eq.~\eqref{eq::local_spectrum} in the main text.
Furthermore, we can define a sublattice spectral function similar as in Eq.~\eqref{eq::arpes}, but restrict the hole creation/ annihilation operator to one sublattice, Eq.~\eqref{eq::FourierSub}. Although this spectrum can not be measured experimentally, it can be computed numerically and yields additional insights, as demonstrated for the Kitaev-Heisenberg model~\cite{Trousselet2014}. In this sublattice spectral function, the lower branch of the free-hole spectrum has finite spectral weight. By contrast, in the experimentally accessible hole spectrum Fig.~\ref{fig::arpesFM}, the lower branch remains dark due to cancellations from phase factors in Eq.~\eqref{eq::FourierBZ}.

To improve the finite time restrictions due to limited entanglement with a fixed bond dimension, we use linear prediction~\cite{White2008} combined with a Gaussian envelope after taking the spatial Fourier transform but before the temporal Fourier transform to obtain the spectral function, Eq.~\eqref{eq::arpes} in the main text. On the other hand, for the local spectrum, Eq.~\eqref{eq::local_spectrum}, we use $i=j$ and sum over the contributions from the two sublattices without the spatial Fourier transformation but still keeping the linear prediction and Gaussian envelope.

To check convergence with bond dimension $\chi$, we compute the time evolution with several values of $\chi \in \{300, 500, 1000\}$. As seen in Fig.~\ref{fig::bond_dimension}, some qualitative features already develop correctly for a small bond dimension $\chi=300$. However, the spectral weight still changes when increasing $\chi$. From $\chi=500$ to $\chi=1000$, the spectral function remains unchanged, indicating convergence. Data shown in all other figures are obtained for bond dimensions $\chi = 500$.\\

\noindent \textbf{Additional data for spectral functions}

\noindent We present additional spectra for different parameters. First, we investigate the spectra for different hopping parameters $t$ when considering ferromagnetic ($J>0$) and antiferromagnetic \mbox{($J<0$)} coupling constants. The resulting spectra are shown in Fig.~\ref{fig::spectra}.

For the ferromagnetic case, Fig.~\ref{fig::spectra}~(a)-(b), we observe only slight changes in the distribution of spectral weight with varying hopping parameters $t$. The general picture remains, further supporting that the spectrum is dominated by free hole hopping. All energy scales are given in units of $t$. Increased hopping leads to a higher contribution from the kinetic energy, which further stabilizes the ferromagnetic order and the associated free coherent hole motion.

Next, we consider the antiferromagnetic spin coupling, $J<0$, and focus on the low-energy scales for different hopping parameters. We find that despite a small redistribution of spectral weight for different $t$, similar features are present throughout the spectra; see Fig.~\ref{fig::spectra}~(c)-(d). We compare the MPS data to the flat bands obtained from the parton construction. The energy difference between the anisotropic flat spinon bands scales with $J_z/J$. Thus, the difference becomes smaller in units of $t$ as we increase the hopping strength. 
Furthermore, for $t/|J|=12$, the flat bands in Fig.~\ref{fig::spectra}~(d) are shifted towards slightly higher energies than predicted from the parton mean-field ansatz. This could indicate an additional interplay between spinons and holons. We note that even for these comparatively large values of the hopping, the effective hopping scale, which is renormalized down by a factor of five to ten, is still in the same order as the exchange. Thus, it is very difficult to numerically reach a true strong coupling limit in which the hole dynamics is effectively much faster than the spin dynamics.

Furthermore, we study the case of isotropic Kitaev couplings $J_z/J=1$ in Fig.~\ref{fig::spectra_iso}. This corresponds to the gapless phase of the Kitaev spin liquid~\cite{Kitaev2006}. A larger MPS bond dimension is needed to faithfully represent even the ground state at half-filling. The subsequent time evolution is also more challenging, and we thus expect small deviations by further increasing the bond dimension beyond the accessible bond dimension of $\chi=1000$, which we used for these simulations. However, qualitatively, we can already see in Fig.~\ref{fig::spectra_iso} that the general behavior in the gapless phase is the same as in the gapped phase: the free-hole response dominates the spectrum for FM Kitaev couplings $J>0$, while in the AFM case, $J<0$ dispersionless features according to the parton picture are dominating. These features emerge from the flat spinon band together with the dominant holon spectral weight at the van Hove singularity. 
Here, we would expect a difference between the gapless and gapped phases. While the gapped phase is characterized by flat spinon bands at different energies along the x-/y-bonds and z-bonds, they should have the same energy for isotropic couplings in the gapless phase. Unfortunately, we do not see this difference clearly in our numerical spectra, which could be due to the finite energy resolution or possible mixing of the spinons and holons when including further interactions. 
Moreover, similar to the gapped case, the flat spinons contribute dominantly to the spectrum, and the dispersive gapless spinon does not possess any significant matrix elements. We also do not see any other contributions in the MPS data that could be directly attributed to dispersive spinons. \\

\noindent \textbf{Parton mean-field theory}

\noindent When we add a hole hopping term $H_t$ to the Kitaev model, the whole Hamiltonian Eq.~\eqref{eq::Kitaev_tJ} is no longer exactly solvable. A general approach to describe spin liquids in the presence of doping is a parton mean-field ansatz. To compute the spectral function using this ansatz, we follow Ref.~\cite{You2012} and consider the splitting of a hole into a charge degree of freedom, the \textit{holon}~$\bd{ }$, and a spin degree of freedom, the \textit{spinon}~$\fd{\sigma}$,
\begin{equation}\label{eq::hole_decomposition}
    \co{i\uparrow} = \frac{1}{\sqrt{2}}(\bd{i1}\fo{i\uparrow}-\bd{i2}\fd{i\downarrow}), \quad \co{i\downarrow} = \frac{1}{\sqrt{2}}(\bd{i1}\fo{i\downarrow}+\bd{i2}\fd{i\uparrow}).
\end{equation}
Here, the holons fulfill the bosonic commutation relations $[\bo{in}, \bd{jm} ]=\delta_{ij}\delta_{nm}$ and the spinons obey fermionic anticommutation relations $\{\fo{i\sigma}, \fd{j\tilde{\sigma}} \} =\delta_{ij}\delta_{\sigma\tilde{\sigma}}$.
By introducing two holon species $\bd{i1}$, $\bd{i2}$ we retain the SU(2) gauge redundancy. At each site we must impose the constraint $\fd{i\uparrow}\fo{i\uparrow} + \fd{i\downarrow}\fo{i\downarrow} + \bd{i1}\bo{i1} - \bd{i2}\bo{i2} = 1$.
The spinons can be related to the Majorana fermion description of the pure Kitaev model Eq.~\eqref{eq::Kitaev_majorana} by fixing a SU(2) gauge~\cite{Burnell2011, Schaffer2012},
\begin{equation}
    \fo{i\uparrow} = \frac{1}{\sqrt{2}} (\chi^0_i + i \chi^z_i), \quad \fo{i\downarrow} = \frac{1}{\sqrt{2}} (i\chi^x_i - \chi^y_i).
\end{equation}
Therefore, we can rewrite the Kitaev model in terms of these spinons:
\begin{equation}\label{eq::Kitaev_spinon}
    H_K = -\sum_{\alpha,\sigma} \sum_{\langle i, j \rangle_\alpha} \left( \tilde{J}^\sigma_\alpha\fd{i, \sigma}\fo{j, \sigma} + \Delta^\sigma_\alpha \fd{i, \sigma}\fd{j, \sigma} + \mathrm{h.c.} \right),
\end{equation}
 where the parameters ${\tilde{J}}$, $\Delta$ are determined self-consistently~\cite{Burnell2011}.

Next, we include the hole hopping term in our description. In our ansatz, we assume that the holon and the spinon are deconfined~\cite{Senthil2008, Podolsky2009} and carry out a mean-field decoupling~\cite{You2012}
\begin{align}\label{eq::hopping_mf}
    H_t = -t \sum_{\langle i,j \rangle, \sigma} &\left( \cd{i\sigma} \co{j\sigma} + \mathrm{h.c.} \right) \nonumber \\
    = -\frac{t}{2} \sum_{\langle i,j \rangle, \sigma} &\Big( \bo{i 1} \bd{j 1} \fd{i\sigma} \fo{j\sigma} -\sigma \bo{i1}\bd{j2} \fd{i\sigma} \fd{j-\sigma} \nonumber \\
    &-\sigma \bo{i2} \bd{j1} \fo{i-\sigma} \fo{j\sigma} + \bo{i2} \bd{j2} \fo{i-\sigma} \fd{j-\sigma} + \mathrm{h.c.} \Big) \nonumber \\
    \approx -\frac{t}{2} \sum_{\langle i,j \rangle, \sigma} &\Big( \langle \bo{i 1} \bd{j 1}\rangle \fd{i\sigma} \fo{j\sigma} + \bo{i 1} \bd{j 1} \langle \fd{i\sigma} \fo{j\sigma} \rangle \nonumber \\
    & -\sigma \langle \bo{i1}\bd{j2}\rangle \fd{i\sigma} \fd{j-\sigma} \nonumber -\sigma \bo{i1}\bd{j2} \langle\fd{i\sigma} \fd{j-\sigma} \rangle \nonumber \\
    &-\sigma \langle \bo{i2} \bd{j1}\rangle \fo{i-\sigma} \fo{j\sigma} -\sigma \bo{i2} \bd{j1} \langle\fo{i-\sigma} \fo{j\sigma} \rangle \nonumber \\
    &+ \langle\bo{i2} \bd{j2}\rangle \fo{i-\sigma} \fd{j-\sigma} + \bo{i2} \bd{j2} \langle\fo{i-\sigma} \fd{j-\sigma}\rangle + \mathrm{h.c.} \Big) \nonumber \\
    = -\frac{t}{2} \sum_{\langle i,j \rangle} &\left(W_{ij} \bo{i1}\bd{j1} - W_{ij}^\ast \bo{i2}\bd{j2} + \mathrm{h.c.} \right),
\end{align}
where we defined $W_{ij}=\sum_\sigma \langle \fd{i\sigma} \fo{j\sigma} \rangle$ and take the expectation value of the undoped Kitaev spin-liquid ground state. Moreover, we used that $\langle \bo{i\mu} \bd{j\mu} \rangle=0$ because after the creation of a holon at site $j$ in the ground state, there cannot be another hole at site $i$ for $i\neq j$.
Furthermore, we have $\langle \fd{i\sigma} \fd{j-\sigma} \rangle=0$ since these pairing terms do not conserve spin parity. Thus, the full Hamiltonian decomposes into $H=H_t+H_K$, where the first part only acts on the holon and the second part describes the spinon dynamics. Accordingly, for the spectral function, we can employ the spectral building principle similar to the one-dimensional $t$-$J$ model~\cite{Bohrdt2018}.
Concretely, the time-dependent correlation function Eq.~\eqref{eq::Cij}, factorizes into a product of the holon and the spinon part.
\begin{align}\label{eq::Cij_parton}
    C_{ij}(\tau)=&\sum_\sigma \Big( \bra{\psi_0} \e^{iH_K\tau} \fd{i\sigma} \e^{-iH_K \tau} \fo{j\sigma} \ket{\psi_0} \bra{0} \e^{iH_t\tau} \bo{i1} \e^{-iH_t \tau} \bd{j1} \ket{0} \nonumber \\
    & +\bra{\psi_0} \e^{iH_K\tau} \fo{i\sigma} \e^{-iH_K \tau} \fd{j\sigma} \ket{\psi_0} \bra{0} \e^{iH_t \tau} \bo{i2} \e^{-iH_t \tau} \bd{j2} \ket{0} \Big)
\end{align}
After Fourier transformation to momentum space, the spectrum is given by the convolution of the individual holon and spinon spectral functions $A_{\mathrm{h}}(\kvec, \omega)$ and $A_{\mathrm{sp}}(\kvec, \omega)$, respectively.
\begin{align}\label{eq::arpes_parton}
    A(\kvec, \omega) &= \int \mathrm{d}\nu \mathrm{d}\qvec \, A_{\mathrm{sp}}(\kvec-\qvec, \omega-\nu) A_{\mathrm{h}}(\qvec, \nu) \nonumber \\
    &= \sum_{\alpha, \beta} \int \mathrm{d}\qvec \,Z_{\mathrm{sp}}^\alpha(\kvec-\qvec) Z_{\mathrm{h}}^\beta(\qvec) \delta(\omega-\varepsilon^{\alpha}_{\mathrm{sp}}(\kvec-\qvec)-\varepsilon^{\beta}_{\mathrm{h}}(\qvec)),
\end{align}
with labels $\alpha$, $\beta$ for the multiple bands of the holon and spinon mean-field Hamiltonians corresponding to different dispersions $\varepsilon(\kvec)$ and quasiparticle weights $Z(\kvec)$. As discussed in the previous section, the periodicity in ARPES experiments extends to a larger BZ because of the two-site unit cell of the honeycomb lattice. In Eq.\eqref{eq::arpes_parton}, the internal integration over $\qvec$ runs only over the first BZ, while $\kvec$ has to be computed for the larger BZ. The dispersion relations are periodic in the 1.BZ, but the quasiparticle weights must be calculated according to Eq.~\eqref{eq::FourierBZ}.

The convolution above generally leads to a broad distribution of spectral weight. For large $t/|J|$, the charge degree of freedom is assumed to have a considerable dispersion, whereas the spins have possible low energy modes. Therefore, typically, we expect that spinons and holons appear at different energy scales and the lower edge of the spectrum is given by the spinon dispersion~\cite{Bohrdt2018, Kadow2022}. However, from Eq.~\eqref{eq::hopping_mf}, we see that the effective hopping constant is strongly renormalized. Concretely, for $J_z/J=2.5$, we find $|W_{ij}|\approx0.23$ on x- and y- bonds and $|W_{ij}|\approx0.36$ on z-bonds. Hence, the spin anisotropy directly results in an anisotropic hopping for the holon. With the prefactor $1/2$ in Eq. \eqref{eq::hopping_mf}, the hopping gets renormalized to quite small effective values. Thus, the spinons and holons no longer have separate energy scales, preventing a clear distinction between the two in the spectrum for reasonable parameters.

In Fig.~\ref{fig::parton_spectrum}, we compare the resulting spectra from the parton construction to the MPS data. The parton mean-field theory only gives predictions for the low-energy part of the spectrum. As can be seen in Fig.~\ref{fig::parton_spectrum}~(b) the convolution of holon and spinon according to Eq.~\eqref{eq::arpes_parton} gives rise to several flat bands. This is expected from the spinon component, where flat bands arise from the bond Majorana fermions. Focusing on the anisotropic case $J_z/J=2.5$, we get distinct bands for the x- and y-bonds (dashed lines) compared to the z-bonds (dotted lines). Importantly, the holon dispersion has a van Hove singularity, at which many states with the same energy will contribute to the spectral function convolution leading to an intense spectral peak. This gives rise to the flat band structure seen in the parton spectrum at energies $\varepsilon_{\mathrm{h}}(\kvec_\mathrm{vH}) + \varepsilon_{\mathrm{sp}}^{x,y,z}$. At energies above the van Hove singularities of the holon, there are dispersing bands with very weak spectral weight, which originate from the convolution of the holon with the dispersive matter Majorana modes.

The MPS spectrum features similar flat bands as well; see Fig.~\ref{fig::parton_spectrum}~(a). The lower one is at the same energy as the parton construction and may be interpreted as the lowest spinon mode plus the van Hove singularity of the holon. However, the spectral weight distribution shows qualitatively different behavior, indicating that there are complex interactions between holon and spinon that must be treated beyond the simple mean-field level.
At larger energies $(-\omega + \mu) \sim 1.0 t$, the MPS spectrum shows dispersive features as well, which have some resemblance of the parton ansatz, especially for the $\Gamma$-$M$ momentum cut. However, the MPS spectra carry much more spectral weight at these energies than the parton predictions.\\

\noindent \textbf{\large Data availability}

All data are available on Zenodo upon reasonable request~\cite{Zenodo}. \\

\noindent \textbf{\large Code availability}

The MPS code for generating the data and code for data analysis are available on Zenodo upon reasonable request~\cite{Zenodo}. \\

\bibliography{main.bib}

\begin{thebibliography}{80}%
\makeatletter
\providecommand \@ifxundefined [1]{%
 \@ifx{#1\undefined}
}%
\providecommand \@ifnum [1]{%
 \ifnum #1\expandafter \@firstoftwo
 \else \expandafter \@secondoftwo
 \fi
}%
\providecommand \@ifx [1]{%
 \ifx #1\expandafter \@firstoftwo
 \else \expandafter \@secondoftwo
 \fi
}%
\providecommand \natexlab [1]{#1}%
\providecommand \enquote  [1]{``#1''}%
\providecommand \bibnamefont  [1]{#1}%
\providecommand \bibfnamefont [1]{#1}%
\providecommand \citenamefont [1]{#1}%
\providecommand \href@noop [0]{\@secondoftwo}%
\providecommand \href [0]{\begingroup \@sanitize@url \@href}%
\providecommand \@href[1]{\@@startlink{#1}\@@href}%
\providecommand \@@href[1]{\endgroup#1\@@endlink}%
\providecommand \@sanitize@url [0]{\catcode `\\12\catcode `\$12\catcode
  `\&12\catcode `\#12\catcode `\^12\catcode `\_12\catcode `\%12\relax}%
\providecommand \@@startlink[1]{}%
\providecommand \@@endlink[0]{}%
\providecommand \url  [0]{\begingroup\@sanitize@url \@url }%
\providecommand \@url [1]{\endgroup\@href {#1}{\urlprefix }}%
\providecommand \urlprefix  [0]{URL }%
\providecommand \Eprint [0]{\href }%
\providecommand \doibase [0]{https://doi.org/}%
\providecommand \selectlanguage [0]{\@gobble}%
\providecommand \bibinfo  [0]{\@secondoftwo}%
\providecommand \bibfield  [0]{\@secondoftwo}%
\providecommand \translation [1]{[#1]}%
\providecommand \BibitemOpen [0]{}%
\providecommand \bibitemStop [0]{}%
\providecommand \bibitemNoStop [0]{.\EOS\space}%
\providecommand \EOS [0]{\spacefactor3000\relax}%
\providecommand \BibitemShut  [1]{\csname bibitem#1\endcsname}%
\let\auto@bib@innerbib\@empty
\bibitem [{\citenamefont {Savary}\ and\ \citenamefont
  {Balents}(2017)}]{Savary2017}%
  \BibitemOpen
  \bibfield  {author} {\bibinfo {author} {\bibfnamefont {L.}~\bibnamefont
  {Savary}}\ and\ \bibinfo {author} {\bibfnamefont {L.}~\bibnamefont
  {Balents}},\ }\bibfield  {title} {\bibinfo {title} {Quantum {Spin}
  {Liquids}},\ }\href {https://doi.org/10.1088/0034-4885/80/1/016502}
  {\bibfield  {journal} {\bibinfo  {journal} {Reports on Progress in Physics}\
  }\textbf {\bibinfo {volume} {80}},\ \bibinfo {pages} {016502} (\bibinfo
  {year} {2017})}\BibitemShut {NoStop}%
\bibitem [{\citenamefont {Balents}(2010)}]{Balents10}%
  \BibitemOpen
  \bibfield  {author} {\bibinfo {author} {\bibfnamefont {L.}~\bibnamefont
  {Balents}},\ }\bibfield  {title} {\bibinfo {title} {Spin liquids in
  frustrated magnets},\ }\href {http://dx.doi.org/10.1038/nature08917}
  {\bibfield  {journal} {\bibinfo  {journal} {Nature}\ }\textbf {\bibinfo
  {volume} {464}},\ \bibinfo {pages} {199} (\bibinfo {year}
  {2010})}\BibitemShut {NoStop}%
\bibitem [{\citenamefont {Zhou}\ \emph {et~al.}(2017)\citenamefont {Zhou},
  \citenamefont {Kanoda},\ and\ \citenamefont {Ng}}]{Zhou2017}%
  \BibitemOpen
  \bibfield  {author} {\bibinfo {author} {\bibfnamefont {Y.}~\bibnamefont
  {Zhou}}, \bibinfo {author} {\bibfnamefont {K.}~\bibnamefont {Kanoda}},\ and\
  \bibinfo {author} {\bibfnamefont {T.-K.}\ \bibnamefont {Ng}},\ }\bibfield
  {title} {\bibinfo {title} {Quantum spin liquid states},\ }\href
  {https://doi.org/10.1103/RevModPhys.89.025003} {\bibfield  {journal}
  {\bibinfo  {journal} {Rev. Mod. Phys.}\ }\textbf {\bibinfo {volume} {89}},\
  \bibinfo {pages} {025003} (\bibinfo {year} {2017})}\BibitemShut {NoStop}%
\bibitem [{\citenamefont {Knolle}\ and\ \citenamefont
  {Moessner}(2019)}]{Knolle2019}%
  \BibitemOpen
  \bibfield  {author} {\bibinfo {author} {\bibfnamefont {J.}~\bibnamefont
  {Knolle}}\ and\ \bibinfo {author} {\bibfnamefont {R.}~\bibnamefont
  {Moessner}},\ }\bibfield  {title} {\bibinfo {title} {A field guide to spin
  liquids},\ }\href {https://doi.org/10.1146/annurev-conmatphys-031218-013401}
  {\bibfield  {journal} {\bibinfo  {journal} {Annu. Rev. Condens. Matter
  Phys.}\ }\textbf {\bibinfo {volume} {10}},\ \bibinfo {pages} {451} (\bibinfo
  {year} {2019})}\BibitemShut {NoStop}%
\bibitem [{\citenamefont {Broholm}\ \emph {et~al.}(2020)\citenamefont {Broholm}
  \emph {et~al.}}]{Broholm2020}%
  \BibitemOpen
  \bibfield  {author} {\bibinfo {author} {\bibfnamefont {C.}~\bibnamefont
  {Broholm}} \emph {et~al.},\ }\bibfield  {title} {\bibinfo {title} {Quantum
  spin liquids},\ }\href
  {https://science.sciencemag.org/content/367/6475/eaay0668} {\bibfield
  {journal} {\bibinfo  {journal} {Science}\ }\textbf {\bibinfo {volume} {367}}
  (\bibinfo {year} {2020})}\BibitemShut {NoStop}%
\bibitem [{\citenamefont {Wen}(1991)}]{Wen1991}%
  \BibitemOpen
  \bibfield  {author} {\bibinfo {author} {\bibfnamefont {X.~G.}\ \bibnamefont
  {Wen}},\ }\bibfield  {title} {\bibinfo {title} {Mean-field theory of
  spin-liquid states with finite energy gap and topological orders},\ }\href
  {https://doi.org/10.1103/PhysRevB.44.2664} {\bibfield  {journal} {\bibinfo
  {journal} {Phys. Rev. B}\ }\textbf {\bibinfo {volume} {44}},\ \bibinfo
  {pages} {2664} (\bibinfo {year} {1991})}\BibitemShut {NoStop}%
\bibitem [{\citenamefont {Wen}(2002)}]{Wen2002}%
  \BibitemOpen
  \bibfield  {author} {\bibinfo {author} {\bibfnamefont {X.-G.}\ \bibnamefont
  {Wen}},\ }\bibfield  {title} {\bibinfo {title} {Quantum orders and symmetric
  spin liquids},\ }\href {https://doi.org/10.1103/PhysRevB.65.165113}
  {\bibfield  {journal} {\bibinfo  {journal} {Phys. Rev. B}\ }\textbf {\bibinfo
  {volume} {65}},\ \bibinfo {pages} {165113} (\bibinfo {year}
  {2002})}\BibitemShut {NoStop}%
\bibitem [{\citenamefont {Kitaev}(2006)}]{Kitaev2006}%
  \BibitemOpen
  \bibfield  {author} {\bibinfo {author} {\bibfnamefont {A.}~\bibnamefont
  {Kitaev}},\ }\bibfield  {title} {\bibinfo {title} {Anyons in an exactly
  solved model and beyond},\ }\href {https://doi.org/10.1016/j.aop.2005.10.005}
  {\bibfield  {journal} {\bibinfo  {journal} {Annals of Physics}\ }\textbf
  {\bibinfo {volume} {321}},\ \bibinfo {pages} {2} (\bibinfo {year}
  {2006})}\BibitemShut {NoStop}%
\bibitem [{\citenamefont {Jackeli}\ and\ \citenamefont
  {Khaliullin}(2009)}]{Jackeli2009}%
  \BibitemOpen
  \bibfield  {author} {\bibinfo {author} {\bibfnamefont {G.}~\bibnamefont
  {Jackeli}}\ and\ \bibinfo {author} {\bibfnamefont {G.}~\bibnamefont
  {Khaliullin}},\ }\bibfield  {title} {\bibinfo {title} {Mott {Insulators} in
  the {Strong} {Spin}-{Orbit} {Coupling} {Limit}: {From} {Heisenberg} to a
  {Quantum} {Compass} and {Kitaev} {Models}},\ }\href
  {https://doi.org/10.1103/PhysRevLett.102.017205} {\bibfield  {journal}
  {\bibinfo  {journal} {Phys. Rev. Lett.}\ }\textbf {\bibinfo {volume} {102}},\
  \bibinfo {pages} {017205} (\bibinfo {year} {2009})}\BibitemShut {NoStop}%
\bibitem [{\citenamefont {Rau}\ \emph {et~al.}(2014)\citenamefont {Rau},
  \citenamefont {Lee},\ and\ \citenamefont {Kee}}]{Rau2014}%
  \BibitemOpen
  \bibfield  {author} {\bibinfo {author} {\bibfnamefont {J.~G.}\ \bibnamefont
  {Rau}}, \bibinfo {author} {\bibfnamefont {E.~K.-H.}\ \bibnamefont {Lee}},\
  and\ \bibinfo {author} {\bibfnamefont {H.-Y.}\ \bibnamefont {Kee}},\
  }\bibfield  {title} {\bibinfo {title} {Generic {Spin} {Model} for the
  {Honeycomb} {Iridates} beyond the {Kitaev} {Limit}},\ }\href
  {https://doi.org/10.1103/PhysRevLett.112.077204} {\bibfield  {journal}
  {\bibinfo  {journal} {Phys. Rev. Lett.}\ }\textbf {\bibinfo {volume} {112}},\
  \bibinfo {pages} {077204} (\bibinfo {year} {2014})}\BibitemShut {NoStop}%
\bibitem [{\citenamefont {Banerjee}\ \emph {et~al.}(2016)\citenamefont
  {Banerjee} \emph {et~al.}}]{Banerjee2016}%
  \BibitemOpen
  \bibfield  {author} {\bibinfo {author} {\bibfnamefont {A.}~\bibnamefont
  {Banerjee}} \emph {et~al.},\ }\bibfield  {title} {\bibinfo {title} {Proximate
  {Kitaev} quantum spin liquid behaviour in a honeycomb magnet},\ }\href
  {https://doi.org/10.1038/nmat4604} {\bibfield  {journal} {\bibinfo  {journal}
  {Nature Materials}\ }\textbf {\bibinfo {volume} {15}},\ \bibinfo {pages}
  {733} (\bibinfo {year} {2016})}\BibitemShut {NoStop}%
\bibitem [{\citenamefont {Takagi}\ \emph {et~al.}(2019)\citenamefont {Takagi},
  \citenamefont {Takayama}, \citenamefont {Jackeli}, \citenamefont
  {Khaliullin},\ and\ \citenamefont {Nagler}}]{Takagi2019}%
  \BibitemOpen
  \bibfield  {author} {\bibinfo {author} {\bibfnamefont {H.}~\bibnamefont
  {Takagi}}, \bibinfo {author} {\bibfnamefont {T.}~\bibnamefont {Takayama}},
  \bibinfo {author} {\bibfnamefont {G.}~\bibnamefont {Jackeli}}, \bibinfo
  {author} {\bibfnamefont {G.}~\bibnamefont {Khaliullin}},\ and\ \bibinfo
  {author} {\bibfnamefont {S.~E.}\ \bibnamefont {Nagler}},\ }\bibfield  {title}
  {\bibinfo {title} {Concept and realization of {Kitaev} quantum spin
  liquids},\ }\href {https://doi.org/10.1038/s42254-019-0038-2} {\bibfield
  {journal} {\bibinfo  {journal} {Nature Reviews Physics}\ }\textbf {\bibinfo
  {volume} {1}},\ \bibinfo {pages} {264} (\bibinfo {year} {2019})}\BibitemShut
  {NoStop}%
\bibitem [{\citenamefont {Chaloupka}\ \emph {et~al.}(2010)\citenamefont
  {Chaloupka}, \citenamefont {Jackeli},\ and\ \citenamefont
  {Khaliullin}}]{Chaloupka2010}%
  \BibitemOpen
  \bibfield  {author} {\bibinfo {author} {\bibfnamefont {J.}~\bibnamefont
  {Chaloupka}}, \bibinfo {author} {\bibfnamefont {G.}~\bibnamefont {Jackeli}},\
  and\ \bibinfo {author} {\bibfnamefont {G.}~\bibnamefont {Khaliullin}},\
  }\bibfield  {title} {\bibinfo {title} {Kitaev-{Heisenberg} {Model} on a
  {Honeycomb} {Lattice}: {Possible} {Exotic} {Phases} in {Iridium} {Oxides}
  ${A}_2\mathrm{IrO}_3$},\ }\href
  {https://doi.org/10.1103/PhysRevLett.105.027204} {\bibfield  {journal}
  {\bibinfo  {journal} {Phys. Rev. Lett.}\ }\textbf {\bibinfo {volume} {105}},\
  \bibinfo {pages} {027204} (\bibinfo {year} {2010})}\BibitemShut {NoStop}%
\bibitem [{\citenamefont {Schaffer}\ \emph {et~al.}(2012)\citenamefont
  {Schaffer}, \citenamefont {Bhattacharjee},\ and\ \citenamefont
  {Kim}}]{Schaffer2012}%
  \BibitemOpen
  \bibfield  {author} {\bibinfo {author} {\bibfnamefont {R.}~\bibnamefont
  {Schaffer}}, \bibinfo {author} {\bibfnamefont {S.}~\bibnamefont
  {Bhattacharjee}},\ and\ \bibinfo {author} {\bibfnamefont {Y.~B.}\
  \bibnamefont {Kim}},\ }\bibfield  {title} {\bibinfo {title} {Quantum phase
  transition in {Heisenberg}-{Kitaev} model},\ }\href
  {https://doi.org/10.1103/PhysRevB.86.224417} {\bibfield  {journal} {\bibinfo
  {journal} {Phys. Rev. B}\ }\textbf {\bibinfo {volume} {86}},\ \bibinfo
  {pages} {224417} (\bibinfo {year} {2012})}\BibitemShut {NoStop}%
\bibitem [{\citenamefont {Osorio~Iregui}\ \emph {et~al.}(2014)\citenamefont
  {Osorio~Iregui}, \citenamefont {Corboz},\ and\ \citenamefont
  {Troyer}}]{Iregui2014}%
  \BibitemOpen
  \bibfield  {author} {\bibinfo {author} {\bibfnamefont {J.}~\bibnamefont
  {Osorio~Iregui}}, \bibinfo {author} {\bibfnamefont {P.}~\bibnamefont
  {Corboz}},\ and\ \bibinfo {author} {\bibfnamefont {M.}~\bibnamefont
  {Troyer}},\ }\bibfield  {title} {\bibinfo {title} {Probing the stability of
  the spin-liquid phases in the kitaev-heisenberg model using tensor network
  algorithms},\ }\href {https://doi.org/10.1103/PhysRevB.90.195102} {\bibfield
  {journal} {\bibinfo  {journal} {Phys. Rev. B}\ }\textbf {\bibinfo {volume}
  {90}},\ \bibinfo {pages} {195102} (\bibinfo {year} {2014})}\BibitemShut
  {NoStop}%
\bibitem [{\citenamefont {Gohlke}\ \emph {et~al.}(2017)\citenamefont {Gohlke},
  \citenamefont {Verresen}, \citenamefont {Moessner},\ and\ \citenamefont
  {Pollmann}}]{Gohlke2017}%
  \BibitemOpen
  \bibfield  {author} {\bibinfo {author} {\bibfnamefont {M.}~\bibnamefont
  {Gohlke}}, \bibinfo {author} {\bibfnamefont {R.}~\bibnamefont {Verresen}},
  \bibinfo {author} {\bibfnamefont {R.}~\bibnamefont {Moessner}},\ and\
  \bibinfo {author} {\bibfnamefont {F.}~\bibnamefont {Pollmann}},\ }\bibfield
  {title} {\bibinfo {title} {Dynamics of the kitaev-heisenberg model},\ }\href
  {https://doi.org/10.1103/physrevlett.119.157203} {\bibfield  {journal}
  {\bibinfo  {journal} {Phys. Rev. Lett.}\ }\textbf {\bibinfo {volume} {119}},\
  \bibinfo {pages} {157203} (\bibinfo {year} {2017})}\BibitemShut {NoStop}%
\bibitem [{\citenamefont {Halász}\ \emph {et~al.}(2014)\citenamefont
  {Halász}, \citenamefont {Chalker},\ and\ \citenamefont
  {Moessner}}]{Halasz2014}%
  \BibitemOpen
  \bibfield  {author} {\bibinfo {author} {\bibfnamefont {G.~B.}\ \bibnamefont
  {Halász}}, \bibinfo {author} {\bibfnamefont {J.~T.}\ \bibnamefont
  {Chalker}},\ and\ \bibinfo {author} {\bibfnamefont {R.}~\bibnamefont
  {Moessner}},\ }\bibfield  {title} {\bibinfo {title} {Doping a topological
  quantum spin liquid: slow holes in the {Kitaev} honeycomb model},\ }\href
  {https://doi.org/10.1103/PhysRevB.90.035145} {\bibfield  {journal} {\bibinfo
  {journal} {Phys. Rev. B}\ }\textbf {\bibinfo {volume} {90}},\ \bibinfo
  {pages} {035145} (\bibinfo {year} {2014})}\BibitemShut {NoStop}%
\bibitem [{\citenamefont {Halász}\ and\ \citenamefont
  {Chalker}(2016)}]{Halasz2016}%
  \BibitemOpen
  \bibfield  {author} {\bibinfo {author} {\bibfnamefont {G.~B.}\ \bibnamefont
  {Halász}}\ and\ \bibinfo {author} {\bibfnamefont {J.~T.}\ \bibnamefont
  {Chalker}},\ }\bibfield  {title} {\bibinfo {title} {Coherent hole propagation
  in an exactly solvable gapless spin liquid},\ }\href
  {https://doi.org/10.1103/PhysRevB.94.235105} {\bibfield  {journal} {\bibinfo
  {journal} {Phys. Rev. B}\ }\textbf {\bibinfo {volume} {94}},\ \bibinfo
  {pages} {235105} (\bibinfo {year} {2016})}\BibitemShut {NoStop}%
\bibitem [{\citenamefont {You}\ \emph {et~al.}(2012)\citenamefont {You},
  \citenamefont {Kimchi},\ and\ \citenamefont {Vishwanath}}]{You2012}%
  \BibitemOpen
  \bibfield  {author} {\bibinfo {author} {\bibfnamefont {Y.-Z.}\ \bibnamefont
  {You}}, \bibinfo {author} {\bibfnamefont {I.}~\bibnamefont {Kimchi}},\ and\
  \bibinfo {author} {\bibfnamefont {A.}~\bibnamefont {Vishwanath}},\ }\bibfield
   {title} {\bibinfo {title} {Doping a spin-orbit {Mott} insulator:
  {Topological} superconductivity from the {Kitaev}-{Heisenberg} model and
  possible application to ({Na}$_2$/{Li}$_2$){IrO}$_3$},\ }\href
  {https://doi.org/10.1103/PhysRevB.86.085145} {\bibfield  {journal} {\bibinfo
  {journal} {Phys. Rev. B}\ }\textbf {\bibinfo {volume} {86}},\ \bibinfo
  {pages} {085145} (\bibinfo {year} {2012})}\BibitemShut {NoStop}%
\bibitem [{\citenamefont {Hyart}\ \emph {et~al.}(2012)\citenamefont {Hyart},
  \citenamefont {Wright}, \citenamefont {Khaliullin},\ and\ \citenamefont
  {Rosenow}}]{Hyart2012}%
  \BibitemOpen
  \bibfield  {author} {\bibinfo {author} {\bibfnamefont {T.}~\bibnamefont
  {Hyart}}, \bibinfo {author} {\bibfnamefont {A.~R.}\ \bibnamefont {Wright}},
  \bibinfo {author} {\bibfnamefont {G.}~\bibnamefont {Khaliullin}},\ and\
  \bibinfo {author} {\bibfnamefont {B.}~\bibnamefont {Rosenow}},\ }\bibfield
  {title} {\bibinfo {title} {Competition between $d$-wave and topological
  $p$-wave superconducting phases in the doped {Kitaev}-{Heisenberg} model},\
  }\href {https://doi.org/10.1103/PhysRevB.85.140510} {\bibfield  {journal}
  {\bibinfo  {journal} {Phys. Rev. B}\ }\textbf {\bibinfo {volume} {85}},\
  \bibinfo {pages} {140510} (\bibinfo {year} {2012})}\BibitemShut {NoStop}%
\bibitem [{\citenamefont {Okamoto}(2013)}]{Okamoto2013}%
  \BibitemOpen
  \bibfield  {author} {\bibinfo {author} {\bibfnamefont {S.}~\bibnamefont
  {Okamoto}},\ }\bibfield  {title} {\bibinfo {title} {Global phase diagram of a
  doped {Kitaev}-{Heisenberg} model},\ }\href
  {https://doi.org/10.1103/PhysRevB.87.064508} {\bibfield  {journal} {\bibinfo
  {journal} {Phys. Rev. B}\ }\textbf {\bibinfo {volume} {87}},\ \bibinfo
  {pages} {064508} (\bibinfo {year} {2013})}\BibitemShut {NoStop}%
\bibitem [{\citenamefont {Scherer}\ \emph {et~al.}(2014)\citenamefont
  {Scherer}, \citenamefont {Scherer}, \citenamefont {Khaliullin}, \citenamefont
  {Honerkamp},\ and\ \citenamefont {Rosenow}}]{Scherer2014}%
  \BibitemOpen
  \bibfield  {author} {\bibinfo {author} {\bibfnamefont {D.~D.}\ \bibnamefont
  {Scherer}}, \bibinfo {author} {\bibfnamefont {M.~M.}\ \bibnamefont
  {Scherer}}, \bibinfo {author} {\bibfnamefont {G.}~\bibnamefont {Khaliullin}},
  \bibinfo {author} {\bibfnamefont {C.}~\bibnamefont {Honerkamp}},\ and\
  \bibinfo {author} {\bibfnamefont {B.}~\bibnamefont {Rosenow}},\ }\bibfield
  {title} {\bibinfo {title} {Unconventional pairing and electronic dimerization
  instabilities in the doped {Kitaev}-{Heisenberg} model},\ }\href
  {https://doi.org/10.1103/PhysRevB.90.045135} {\bibfield  {journal} {\bibinfo
  {journal} {Phys. Rev. B}\ }\textbf {\bibinfo {volume} {90}},\ \bibinfo
  {pages} {045135} (\bibinfo {year} {2014})}\BibitemShut {NoStop}%
\bibitem [{\citenamefont {Mei}(2012)}]{Mei2012}%
  \BibitemOpen
  \bibfield  {author} {\bibinfo {author} {\bibfnamefont {J.-W.}\ \bibnamefont
  {Mei}},\ }\bibfield  {title} {\bibinfo {title} {Possible {F}ermi {L}iquid in
  the {L}ightly {D}oped {K}itaev {S}pin {L}iquid},\ }\href
  {https://doi.org/10.1103/PhysRevLett.108.227207} {\bibfield  {journal}
  {\bibinfo  {journal} {Phys. Rev. Lett.}\ }\textbf {\bibinfo {volume} {108}},\
  \bibinfo {pages} {227207} (\bibinfo {year} {2012})}\BibitemShut {NoStop}%
\bibitem [{\citenamefont {Meden}\ and\ \citenamefont
  {Sch\"onhammer}(1992)}]{Meden1992}%
  \BibitemOpen
  \bibfield  {author} {\bibinfo {author} {\bibfnamefont {V.}~\bibnamefont
  {Meden}}\ and\ \bibinfo {author} {\bibfnamefont {K.}~\bibnamefont
  {Sch\"onhammer}},\ }\bibfield  {title} {\bibinfo {title} {Spectral functions
  for the tomonaga-luttinger model},\ }\href
  {https://doi.org/10.1103/PhysRevB.46.15753} {\bibfield  {journal} {\bibinfo
  {journal} {Phys. Rev. B}\ }\textbf {\bibinfo {volume} {46}},\ \bibinfo
  {pages} {15753} (\bibinfo {year} {1992})}\BibitemShut {NoStop}%
\bibitem [{\citenamefont {Voit}(1993)}]{Voit1993}%
  \BibitemOpen
  \bibfield  {author} {\bibinfo {author} {\bibfnamefont {J.}~\bibnamefont
  {Voit}},\ }\bibfield  {title} {\bibinfo {title} {Charge-spin separation and
  the spectral properties of luttinger liquids},\ }\href
  {https://doi.org/10.1103/PhysRevB.47.6740} {\bibfield  {journal} {\bibinfo
  {journal} {Phys. Rev. B}\ }\textbf {\bibinfo {volume} {47}},\ \bibinfo
  {pages} {6740} (\bibinfo {year} {1993})}\BibitemShut {NoStop}%
\bibitem [{\citenamefont {Giamarchi}(2004)}]{Giamarchi2004}%
  \BibitemOpen
  \bibfield  {author} {\bibinfo {author} {\bibfnamefont {T.}~\bibnamefont
  {Giamarchi}},\ }\href {https://books.google.de/books?id=1MwTDAAAQBAJ} {\emph
  {\bibinfo {title} {Quantum Physics in One Dimension}}},\ International Series
  of Monographs on Physics\ (\bibinfo  {publisher} {Clarendon Press},\ \bibinfo
  {year} {2004})\BibitemShut {NoStop}%
\bibitem [{\citenamefont {Senthil}(2008)}]{Senthil2008}%
  \BibitemOpen
  \bibfield  {author} {\bibinfo {author} {\bibfnamefont {T.}~\bibnamefont
  {Senthil}},\ }\bibfield  {title} {\bibinfo {title} {Theory of a continuous
  {M}ott transition in two dimensions},\ }\href
  {https://doi.org/10.1103/PhysRevB.78.045109} {\bibfield  {journal} {\bibinfo
  {journal} {Phys. Rev. B}\ }\textbf {\bibinfo {volume} {78}},\ \bibinfo
  {pages} {045109} (\bibinfo {year} {2008})}\BibitemShut {NoStop}%
\bibitem [{\citenamefont {Podolsky}\ \emph {et~al.}(2009)\citenamefont
  {Podolsky}, \citenamefont {Paramekanti}, \citenamefont {Kim},\ and\
  \citenamefont {Senthil}}]{Podolsky2009}%
  \BibitemOpen
  \bibfield  {author} {\bibinfo {author} {\bibfnamefont {D.}~\bibnamefont
  {Podolsky}}, \bibinfo {author} {\bibfnamefont {A.}~\bibnamefont
  {Paramekanti}}, \bibinfo {author} {\bibfnamefont {Y.~B.}\ \bibnamefont
  {Kim}},\ and\ \bibinfo {author} {\bibfnamefont {T.}~\bibnamefont {Senthil}},\
  }\bibfield  {title} {\bibinfo {title} {Mott transition between a spin-liquid
  insulator and a metal in three dimensions},\ }\href
  {https://doi.org/10.1103/PhysRevLett.102.186401} {\bibfield  {journal}
  {\bibinfo  {journal} {Phys. Rev. Lett.}\ }\textbf {\bibinfo {volume} {102}},\
  \bibinfo {pages} {186401} (\bibinfo {year} {2009})}\BibitemShut {NoStop}%
\bibitem [{\citenamefont {Läuchli}\ and\ \citenamefont
  {Poilblanc}(2004)}]{Laeuchli2004}%
  \BibitemOpen
  \bibfield  {author} {\bibinfo {author} {\bibfnamefont {A.}~\bibnamefont
  {Läuchli}}\ and\ \bibinfo {author} {\bibfnamefont {D.}~\bibnamefont
  {Poilblanc}},\ }\bibfield  {title} {\bibinfo {title} {Spin-{Charge}
  {Separation} in {Two}-{Dimensional} {Frustrated} {Quantum} {Magnets}},\
  }\href {https://doi.org/10.1103/PhysRevLett.92.236404} {\bibfield  {journal}
  {\bibinfo  {journal} {Phys. Rev. Lett.}\ }\textbf {\bibinfo {volume} {92}},\
  \bibinfo {pages} {236404} (\bibinfo {year} {2004})}\BibitemShut {NoStop}%
\bibitem [{\citenamefont {Kadow}\ \emph {et~al.}(2022)\citenamefont {Kadow},
  \citenamefont {Vanderstraeten},\ and\ \citenamefont {Knap}}]{Kadow2022}%
  \BibitemOpen
  \bibfield  {author} {\bibinfo {author} {\bibfnamefont {W.}~\bibnamefont
  {Kadow}}, \bibinfo {author} {\bibfnamefont {L.}~\bibnamefont
  {Vanderstraeten}},\ and\ \bibinfo {author} {\bibfnamefont {M.}~\bibnamefont
  {Knap}},\ }\bibfield  {title} {\bibinfo {title} {Hole spectral function of a
  chiral spin liquid in the triangular lattice {Hubbard} model},\ }\href
  {https://doi.org/10.1103/PhysRevB.106.094417} {\bibfield  {journal} {\bibinfo
   {journal} {Phys. Rev. B}\ }\textbf {\bibinfo {volume} {106}},\ \bibinfo
  {pages} {094417} (\bibinfo {year} {2022})}\BibitemShut {NoStop}%
\bibitem [{\citenamefont {Dagotto}\ \emph {et~al.}(1990)\citenamefont
  {Dagotto}, \citenamefont {Joynt}, \citenamefont {Moreo}, \citenamefont
  {Bacci},\ and\ \citenamefont {Gagliano}}]{Dagotto1990}%
  \BibitemOpen
  \bibfield  {author} {\bibinfo {author} {\bibfnamefont {E.}~\bibnamefont
  {Dagotto}}, \bibinfo {author} {\bibfnamefont {R.}~\bibnamefont {Joynt}},
  \bibinfo {author} {\bibfnamefont {A.}~\bibnamefont {Moreo}}, \bibinfo
  {author} {\bibfnamefont {S.}~\bibnamefont {Bacci}},\ and\ \bibinfo {author}
  {\bibfnamefont {E.}~\bibnamefont {Gagliano}},\ }\bibfield  {title} {\bibinfo
  {title} {Strongly correlated electronic systems with one hole: Dynamical
  properties},\ }\href {https://doi.org/10.1103/PhysRevB.41.9049} {\bibfield
  {journal} {\bibinfo  {journal} {Phys. Rev. B}\ }\textbf {\bibinfo {volume}
  {41}},\ \bibinfo {pages} {9049} (\bibinfo {year} {1990})}\BibitemShut
  {NoStop}%
\bibitem [{\citenamefont {Martinez}\ and\ \citenamefont
  {Horsch}(1991)}]{Martinez1991}%
  \BibitemOpen
  \bibfield  {author} {\bibinfo {author} {\bibfnamefont {G.}~\bibnamefont
  {Martinez}}\ and\ \bibinfo {author} {\bibfnamefont {P.}~\bibnamefont
  {Horsch}},\ }\bibfield  {title} {\bibinfo {title} {Spin polarons in the {t-J}
  model},\ }\href {https://doi.org/10.1103/PhysRevB.44.317} {\bibfield
  {journal} {\bibinfo  {journal} {Phys. Rev. B}\ }\textbf {\bibinfo {volume}
  {44}},\ \bibinfo {pages} {317} (\bibinfo {year} {1991})}\BibitemShut
  {NoStop}%
\bibitem [{\citenamefont {Auerbach}\ and\ \citenamefont
  {Larson}(1991)}]{Auerbach1991}%
  \BibitemOpen
  \bibfield  {author} {\bibinfo {author} {\bibfnamefont {A.}~\bibnamefont
  {Auerbach}}\ and\ \bibinfo {author} {\bibfnamefont {B.~E.}\ \bibnamefont
  {Larson}},\ }\bibfield  {title} {\bibinfo {title} {Small-polaron theory of
  doped antiferromagnets},\ }\href
  {https://doi.org/10.1103/PhysRevLett.66.2262} {\bibfield  {journal} {\bibinfo
   {journal} {Phys. Rev. Lett.}\ }\textbf {\bibinfo {volume} {66}},\ \bibinfo
  {pages} {2262} (\bibinfo {year} {1991})}\BibitemShut {NoStop}%
\bibitem [{\citenamefont {Béran}\ \emph {et~al.}(1996)\citenamefont {Béran},
  \citenamefont {Poilblanc},\ and\ \citenamefont {Laughlin}}]{Beran1996}%
  \BibitemOpen
  \bibfield  {author} {\bibinfo {author} {\bibfnamefont {P.}~\bibnamefont
  {Béran}}, \bibinfo {author} {\bibfnamefont {D.}~\bibnamefont {Poilblanc}},\
  and\ \bibinfo {author} {\bibfnamefont {R.}~\bibnamefont {Laughlin}},\
  }\bibfield  {title} {\bibinfo {title} {Evidence for composite nature of
  quasiparticles in the {2D} {t-J} model},\ }\href
  {https://doi.org/https://doi.org/10.1016/0550-3213(96)00196-4} {\bibfield
  {journal} {\bibinfo  {journal} {Nuclear Physics B}\ }\textbf {\bibinfo
  {volume} {473}},\ \bibinfo {pages} {707} (\bibinfo {year}
  {1996})}\BibitemShut {NoStop}%
\bibitem [{\citenamefont {Laughlin}(1997)}]{Laughlin1997}%
  \BibitemOpen
  \bibfield  {author} {\bibinfo {author} {\bibfnamefont {R.~B.}\ \bibnamefont
  {Laughlin}},\ }\bibfield  {title} {\bibinfo {title} {Evidence for
  quasiparticle decay in photoemission from underdoped cuprates},\ }\href
  {https://doi.org/10.1103/PhysRevLett.79.1726} {\bibfield  {journal} {\bibinfo
   {journal} {Phys. Rev. Lett.}\ }\textbf {\bibinfo {volume} {79}},\ \bibinfo
  {pages} {1726} (\bibinfo {year} {1997})}\BibitemShut {NoStop}%
\bibitem [{\citenamefont {Brunner}\ \emph {et~al.}(2000)\citenamefont
  {Brunner}, \citenamefont {Assaad},\ and\ \citenamefont
  {Muramatsu}}]{Brunner2000}%
  \BibitemOpen
  \bibfield  {author} {\bibinfo {author} {\bibfnamefont {M.}~\bibnamefont
  {Brunner}}, \bibinfo {author} {\bibfnamefont {F.~F.}\ \bibnamefont
  {Assaad}},\ and\ \bibinfo {author} {\bibfnamefont {A.}~\bibnamefont
  {Muramatsu}},\ }\bibfield  {title} {\bibinfo {title} {Single-hole dynamics in
  the {$t\ensuremath{-}J$} model on a square lattice},\ }\href
  {https://doi.org/10.1103/PhysRevB.62.15480} {\bibfield  {journal} {\bibinfo
  {journal} {Phys. Rev. B}\ }\textbf {\bibinfo {volume} {62}},\ \bibinfo
  {pages} {15480} (\bibinfo {year} {2000})}\BibitemShut {NoStop}%
\bibitem [{\citenamefont {Mishchenko}\ \emph {et~al.}(2001)\citenamefont
  {Mishchenko}, \citenamefont {Prokof'ev},\ and\ \citenamefont
  {Svistunov}}]{Mishchenko2001}%
  \BibitemOpen
  \bibfield  {author} {\bibinfo {author} {\bibfnamefont {A.~S.}\ \bibnamefont
  {Mishchenko}}, \bibinfo {author} {\bibfnamefont {N.~V.}\ \bibnamefont
  {Prokof'ev}},\ and\ \bibinfo {author} {\bibfnamefont {B.~V.}\ \bibnamefont
  {Svistunov}},\ }\bibfield  {title} {\bibinfo {title} {Single-hole spectral
  function and spin-charge separation in the {$t\ensuremath{-}J$} model},\
  }\href {https://doi.org/10.1103/PhysRevB.64.033101} {\bibfield  {journal}
  {\bibinfo  {journal} {Phys. Rev. B}\ }\textbf {\bibinfo {volume} {64}},\
  \bibinfo {pages} {033101} (\bibinfo {year} {2001})}\BibitemShut {NoStop}%
\bibitem [{\citenamefont {Bohrdt}\ \emph
  {et~al.}(2020{\natexlab{a}})\citenamefont {Bohrdt}, \citenamefont {Grusdt},\
  and\ \citenamefont {Knap}}]{Bohrdt2020a}%
  \BibitemOpen
  \bibfield  {author} {\bibinfo {author} {\bibfnamefont {A.}~\bibnamefont
  {Bohrdt}}, \bibinfo {author} {\bibfnamefont {F.}~\bibnamefont {Grusdt}},\
  and\ \bibinfo {author} {\bibfnamefont {M.}~\bibnamefont {Knap}},\ }\bibfield
  {title} {\bibinfo {title} {Dynamical formation of a magnetic polaron in a
  two-dimensional quantum antiferromagnet},\ }\href
  {https://doi.org/10.1088/1367-2630/abcfee} {\bibfield  {journal} {\bibinfo
  {journal} {New Journal of Physics}\ }\textbf {\bibinfo {volume} {22}},\
  \bibinfo {pages} {123023} (\bibinfo {year} {2020}{\natexlab{a}})}\BibitemShut
  {NoStop}%
\bibitem [{\citenamefont {Bohrdt}\ \emph
  {et~al.}(2020{\natexlab{b}})\citenamefont {Bohrdt}, \citenamefont {Demler},
  \citenamefont {Pollmann}, \citenamefont {Knap},\ and\ \citenamefont
  {Grusdt}}]{Bohrdt2020}%
  \BibitemOpen
  \bibfield  {author} {\bibinfo {author} {\bibfnamefont {A.}~\bibnamefont
  {Bohrdt}}, \bibinfo {author} {\bibfnamefont {E.}~\bibnamefont {Demler}},
  \bibinfo {author} {\bibfnamefont {F.}~\bibnamefont {Pollmann}}, \bibinfo
  {author} {\bibfnamefont {M.}~\bibnamefont {Knap}},\ and\ \bibinfo {author}
  {\bibfnamefont {F.}~\bibnamefont {Grusdt}},\ }\bibfield  {title} {\bibinfo
  {title} {Parton theory of {ARPES} spectra in anti-ferromagnetic {Mott}
  insulators},\ }\href {https://doi.org/10.1103/PhysRevB.102.035139} {\bibfield
   {journal} {\bibinfo  {journal} {Phys. Rev. B}\ }\textbf {\bibinfo {volume}
  {102}},\ \bibinfo {pages} {035139} (\bibinfo {year}
  {2020}{\natexlab{b}})}\BibitemShut {NoStop}%
\bibitem [{\citenamefont {Wrzosek}\ and\ \citenamefont
  {Wohlfeld}(2021)}]{Wrzosek2021}%
  \BibitemOpen
  \bibfield  {author} {\bibinfo {author} {\bibfnamefont {P.}~\bibnamefont
  {Wrzosek}}\ and\ \bibinfo {author} {\bibfnamefont {K.}~\bibnamefont
  {Wohlfeld}},\ }\bibfield  {title} {\bibinfo {title} {Hole in the
  two-dimensional ising antiferromagnet: Origin of the incoherent spectrum},\
  }\href {https://doi.org/10.1103/PhysRevB.103.035113} {\bibfield  {journal}
  {\bibinfo  {journal} {Phys. Rev. B}\ }\textbf {\bibinfo {volume} {103}},\
  \bibinfo {pages} {035113} (\bibinfo {year} {2021})}\BibitemShut {NoStop}%
\bibitem [{\citenamefont {Nagaoka}(1966)}]{Nagaoka1966}%
  \BibitemOpen
  \bibfield  {author} {\bibinfo {author} {\bibfnamefont {Y.}~\bibnamefont
  {Nagaoka}},\ }\bibfield  {title} {\bibinfo {title} {Ferromagnetism in a
  {Narrow}, {Almost} {Half}-{Filled} $s$ {Band}},\ }\href
  {https://doi.org/10.1103/PhysRev.147.392} {\bibfield  {journal} {\bibinfo
  {journal} {Physical Review}\ }\textbf {\bibinfo {volume} {147}},\ \bibinfo
  {pages} {392} (\bibinfo {year} {1966})}\BibitemShut {NoStop}%
\bibitem [{\citenamefont {Tasaki}(1989)}]{Tasaki1989}%
  \BibitemOpen
  \bibfield  {author} {\bibinfo {author} {\bibfnamefont {H.}~\bibnamefont
  {Tasaki}},\ }\bibfield  {title} {\bibinfo {title} {Extension of {Nagaoka}'s
  theorem on the large-{U} {Hubbard} model},\ }\href
  {https://doi.org/10.1103/PhysRevB.40.9192} {\bibfield  {journal} {\bibinfo
  {journal} {Phys. Rev. B}\ }\textbf {\bibinfo {volume} {40}},\ \bibinfo
  {pages} {9192} (\bibinfo {year} {1989})}\BibitemShut {NoStop}%
\bibitem [{\citenamefont {Dehollain}\ \emph {et~al.}(2020)\citenamefont
  {Dehollain} \emph {et~al.}}]{Dehollain2020}%
  \BibitemOpen
  \bibfield  {author} {\bibinfo {author} {\bibfnamefont {J.~P.}\ \bibnamefont
  {Dehollain}} \emph {et~al.},\ }\bibfield  {title} {\bibinfo {title} {Nagaoka
  ferromagnetism observed in a quantum dot plaquette},\ }\href
  {https://doi.org/10.1038/s41586-020-2051-0} {\bibfield  {journal} {\bibinfo
  {journal} {Nature}\ }\textbf {\bibinfo {volume} {579}},\ \bibinfo {pages}
  {528} (\bibinfo {year} {2020})}\BibitemShut {NoStop}%
\bibitem [{\citenamefont {Ciorciaro}\ \emph {et~al.}(2023)\citenamefont
  {Ciorciaro} \emph {et~al.}}]{Ciorciaro2023}%
  \BibitemOpen
  \bibfield  {author} {\bibinfo {author} {\bibfnamefont {L.}~\bibnamefont
  {Ciorciaro}} \emph {et~al.},\ }\bibfield  {title} {\bibinfo {title} {Kinetic
  {Magnetism} in {Triangular} {Moir}{\textbackslash}'e {Materials}},\ }\href
  {https://doi.org/10.1038/s41586-023-06633-0} {\bibfield  {journal} {\bibinfo
  {journal} {Nature}\ }\textbf {\bibinfo {volume} {623}},\ \bibinfo {pages}
  {509} (\bibinfo {year} {2023})}\BibitemShut {NoStop}%
\bibitem [{\citenamefont {Xu}\ \emph {et~al.}(2023)\citenamefont {Xu} \emph
  {et~al.}}]{Xu2023}%
  \BibitemOpen
  \bibfield  {author} {\bibinfo {author} {\bibfnamefont {M.}~\bibnamefont {Xu}}
  \emph {et~al.},\ }\bibfield  {title} {\bibinfo {title} {Frustration- and
  doping-induced magnetism in a {Fermi}-{Hubbard} simulator},\ }\href
  {https://doi.org/10.1038/s41586-023-06280-5} {\bibfield  {journal} {\bibinfo
  {journal} {Nature}\ }\textbf {\bibinfo {volume} {620}},\ \bibinfo {pages}
  {971} (\bibinfo {year} {2023})}\BibitemShut {NoStop}%
\bibitem [{\citenamefont {Lebrat}\ \emph {et~al.}(2023)\citenamefont {Lebrat}
  \emph {et~al.}}]{Lebrat2023}%
  \BibitemOpen
  \bibfield  {author} {\bibinfo {author} {\bibfnamefont {M.}~\bibnamefont
  {Lebrat}} \emph {et~al.},\ }\bibfield  {title} {\bibinfo {title} {Observation
  of {Nagaoka} {Polarons} in a {Fermi}-{Hubbard} {Quantum} {Simulator}},\
  }\href {http://arxiv.org/abs/2308.12269} {\bibfield  {journal} {\bibinfo
  {journal} {arXiv:2308.12269}\ } (\bibinfo {year} {2023})}\BibitemShut
  {NoStop}%
\bibitem [{\citenamefont {Prichard}\ \emph {et~al.}(2023)\citenamefont
  {Prichard} \emph {et~al.}}]{Prichard2023}%
  \BibitemOpen
  \bibfield  {author} {\bibinfo {author} {\bibfnamefont {M.~L.}\ \bibnamefont
  {Prichard}} \emph {et~al.},\ }\bibfield  {title} {\bibinfo {title} {Directly
  imaging spin polarons in a kinetically frustrated {Hubbard} system},\ }\href
  {http://arxiv.org/abs/2308.12951} {\bibfield  {journal} {\bibinfo  {journal}
  {arXiv:2308.12951}\ } (\bibinfo {year} {2023})}\BibitemShut {NoStop}%
\bibitem [{\citenamefont {Carlström}\ \emph {et~al.}(2016)\citenamefont
  {Carlström}, \citenamefont {Prokof’ev},\ and\ \citenamefont
  {Svistunov}}]{Carlstroem2016}%
  \BibitemOpen
  \bibfield  {author} {\bibinfo {author} {\bibfnamefont {J.}~\bibnamefont
  {Carlström}}, \bibinfo {author} {\bibfnamefont {N.}~\bibnamefont
  {Prokof’ev}},\ and\ \bibinfo {author} {\bibfnamefont {B.}~\bibnamefont
  {Svistunov}},\ }\bibfield  {title} {\bibinfo {title} {Quantum {Walk} in
  {Degenerate} {Spin} {Environments}},\ }\href
  {https://doi.org/10.1103/PhysRevLett.116.247202} {\bibfield  {journal}
  {\bibinfo  {journal} {Phys. Rev. Lett.}\ }\textbf {\bibinfo {volume} {116}},\
  \bibinfo {pages} {247202} (\bibinfo {year} {2016})}\BibitemShut {NoStop}%
\bibitem [{\citenamefont {Kanász-Nagy}\ \emph {et~al.}(2017)\citenamefont
  {Kanász-Nagy} \emph {et~al.}}]{KanaszNagy2017}%
  \BibitemOpen
  \bibfield  {author} {\bibinfo {author} {\bibfnamefont {M.}~\bibnamefont
  {Kanász-Nagy}} \emph {et~al.},\ }\bibfield  {title} {\bibinfo {title}
  {Quantum correlations at infinite temperature: {The} dynamical {Nagaoka}
  effect},\ }\href {https://doi.org/10.1103/PhysRevB.96.014303} {\bibfield
  {journal} {\bibinfo  {journal} {Phys. Rev. B}\ }\textbf {\bibinfo {volume}
  {96}},\ \bibinfo {pages} {014303} (\bibinfo {year} {2017})}\BibitemShut
  {NoStop}%
\bibitem [{\citenamefont {Zhou}\ \emph {et~al.}(2016)\citenamefont {Zhou} \emph
  {et~al.}}]{Zhou2016}%
  \BibitemOpen
  \bibfield  {author} {\bibinfo {author} {\bibfnamefont {X.}~\bibnamefont
  {Zhou}} \emph {et~al.},\ }\bibfield  {title} {\bibinfo {title} {{ARPES} study
  of the {Kitaev} {Candidate} $\alpha$-{RuCl}$_3$},\ }\href
  {https://doi.org/10.1103/PhysRevB.94.161106} {\bibfield  {journal} {\bibinfo
  {journal} {Phys. Rev. B}\ }\textbf {\bibinfo {volume} {94}},\ \bibinfo
  {pages} {161106} (\bibinfo {year} {2016})}\BibitemShut {NoStop}%
\bibitem [{\citenamefont {Sinn}\ \emph {et~al.}(2016)\citenamefont {Sinn} \emph
  {et~al.}}]{Sinn2016}%
  \BibitemOpen
  \bibfield  {author} {\bibinfo {author} {\bibfnamefont {S.}~\bibnamefont
  {Sinn}} \emph {et~al.},\ }\bibfield  {title} {\bibinfo {title} {Electronic
  {Structure} of the {Kitaev} {Material} $\alpha$-{RuCl}$_3$ {Probed} by
  {Photoemission} and {Inverse} {Photoemission} {Spectroscopies}},\ }\href
  {https://doi.org/10.1038/srep39544} {\bibfield  {journal} {\bibinfo
  {journal} {Scientific Reports}\ }\textbf {\bibinfo {volume} {6}},\ \bibinfo
  {pages} {39544} (\bibinfo {year} {2016})}\BibitemShut {NoStop}%
\bibitem [{\citenamefont {Comin}\ \emph {et~al.}(2012)\citenamefont {Comin}
  \emph {et~al.}}]{Comin2012}%
  \BibitemOpen
  \bibfield  {author} {\bibinfo {author} {\bibfnamefont {R.}~\bibnamefont
  {Comin}} \emph {et~al.},\ }\bibfield  {title} {\bibinfo {title}
  {$\mathrm{{Na}}_2\mathrm{IrO}_3$ as a {Novel} {Relativistic} {Mott}
  {Insulator} with a 340-{meV} {Gap}},\ }\href
  {https://doi.org/10.1103/PhysRevLett.109.266406} {\bibfield  {journal}
  {\bibinfo  {journal} {Phys. Rev. Lett.}\ }\textbf {\bibinfo {volume} {109}},\
  \bibinfo {pages} {266406} (\bibinfo {year} {2012})}\BibitemShut {NoStop}%
\bibitem [{\citenamefont {Alidoust}\ \emph {et~al.}(2016)\citenamefont
  {Alidoust} \emph {et~al.}}]{Alidoust2016}%
  \BibitemOpen
  \bibfield  {author} {\bibinfo {author} {\bibfnamefont {N.}~\bibnamefont
  {Alidoust}} \emph {et~al.},\ }\bibfield  {title} {\bibinfo {title}
  {Observation of metallic surface states in the strongly correlated
  {Kitaev}-{Heisenberg} candidate $\mathrm{Na}_2\mathrm{IrO}_3$},\ }\href
  {https://doi.org/10.1103/PhysRevB.93.245132} {\bibfield  {journal} {\bibinfo
  {journal} {Phys. Rev. B}\ }\textbf {\bibinfo {volume} {93}},\ \bibinfo
  {pages} {245132} (\bibinfo {year} {2016})}\BibitemShut {NoStop}%
\bibitem [{\citenamefont {Wang}\ \emph {et~al.}(2023)\citenamefont {Wang},
  \citenamefont {Dong}, \citenamefont {Yu},\ and\ \citenamefont
  {Li}}]{Wang2023}%
  \BibitemOpen
  \bibfield  {author} {\bibinfo {author} {\bibfnamefont {W.}~\bibnamefont
  {Wang}}, \bibinfo {author} {\bibfnamefont {Z.-Y.}\ \bibnamefont {Dong}},
  \bibinfo {author} {\bibfnamefont {S.-L.}\ \bibnamefont {Yu}},\ and\ \bibinfo
  {author} {\bibfnamefont {J.-X.}\ \bibnamefont {Li}},\ }\bibfield  {title}
  {\bibinfo {title} {Spectrum of the hole excitation in spin-orbit {Mott}
  insulator {Na}$_2${IrO}$_3$},\ }\href
  {https://doi.org/10.1088/0256-307X/40/8/087101} {\bibfield  {journal}
  {\bibinfo  {journal} {Chinese Physics Letters}\ }\textbf {\bibinfo {volume}
  {40}},\ \bibinfo {pages} {087101} (\bibinfo {year} {2023})}\BibitemShut
  {NoStop}%
\bibitem [{\citenamefont {Trousselet}\ \emph {et~al.}(2014)\citenamefont
  {Trousselet}, \citenamefont {Horsch}, \citenamefont {Oles},\ and\
  \citenamefont {You}}]{Trousselet2014}%
  \BibitemOpen
  \bibfield  {author} {\bibinfo {author} {\bibfnamefont {F.}~\bibnamefont
  {Trousselet}}, \bibinfo {author} {\bibfnamefont {P.}~\bibnamefont {Horsch}},
  \bibinfo {author} {\bibfnamefont {A.~M.}\ \bibnamefont {Oles}},\ and\
  \bibinfo {author} {\bibfnamefont {W.-L.}\ \bibnamefont {You}},\ }\bibfield
  {title} {\bibinfo {title} {Hole propagation in the {Kitaev}-{Heisenberg}
  model: {From} quasiparticles in quantum {Neel} states to non-{Fermi} liquid
  in the {Kitaev} phase},\ }\href {https://doi.org/10.1103/PhysRevB.90.024404}
  {\bibfield  {journal} {\bibinfo  {journal} {Phys. Rev. B}\ }\textbf {\bibinfo
  {volume} {90}},\ \bibinfo {pages} {024404} (\bibinfo {year}
  {2014})}\BibitemShut {NoStop}%
\bibitem [{\citenamefont {Damascelli}\ \emph {et~al.}(2003)\citenamefont
  {Damascelli}, \citenamefont {Hussain},\ and\ \citenamefont
  {Shen}}]{Damascelli2003}%
  \BibitemOpen
  \bibfield  {author} {\bibinfo {author} {\bibfnamefont {A.}~\bibnamefont
  {Damascelli}}, \bibinfo {author} {\bibfnamefont {Z.}~\bibnamefont
  {Hussain}},\ and\ \bibinfo {author} {\bibfnamefont {Z.-X.}\ \bibnamefont
  {Shen}},\ }\bibfield  {title} {\bibinfo {title} {Angle-resolved photoemission
  studies of the cuprate superconductors},\ }\href
  {https://doi.org/10.1103/RevModPhys.75.473} {\bibfield  {journal} {\bibinfo
  {journal} {Reviews of Modern Physics}\ }\textbf {\bibinfo {volume} {75}},\
  \bibinfo {pages} {473} (\bibinfo {year} {2003})}\BibitemShut {NoStop}%
\bibitem [{\citenamefont {Choi}\ \emph {et~al.}(2018)\citenamefont {Choi},
  \citenamefont {Klein}, \citenamefont {Rosch},\ and\ \citenamefont
  {Kim}}]{Choi2018}%
  \BibitemOpen
  \bibfield  {author} {\bibinfo {author} {\bibfnamefont {W.}~\bibnamefont
  {Choi}}, \bibinfo {author} {\bibfnamefont {P.~W.}\ \bibnamefont {Klein}},
  \bibinfo {author} {\bibfnamefont {A.}~\bibnamefont {Rosch}},\ and\ \bibinfo
  {author} {\bibfnamefont {Y.~B.}\ \bibnamefont {Kim}},\ }\bibfield  {title}
  {\bibinfo {title} {Topological superconductivity in {Kondo}-{Kitaev} model},\
  }\href {https://doi.org/10.1103/PhysRevB.98.155123} {\bibfield  {journal}
  {\bibinfo  {journal} {Phys. Rev. B}\ }\textbf {\bibinfo {volume} {98}},\
  \bibinfo {pages} {155123} (\bibinfo {year} {2018})}\BibitemShut {NoStop}%
\bibitem [{\citenamefont {Choi}\ \emph {et~al.}(2012)\citenamefont {Choi} \emph
  {et~al.}}]{Choi2012}%
  \BibitemOpen
  \bibfield  {author} {\bibinfo {author} {\bibfnamefont {S.~K.}\ \bibnamefont
  {Choi}} \emph {et~al.},\ }\bibfield  {title} {\bibinfo {title} {Spin {Waves}
  and {Revised} {Crystal} {Structure} of {Honeycomb} {Iridate}
  $\mathrm{Na}_2\mathrm{IrO}_3$},\ }\href
  {https://doi.org/10.1103/PhysRevLett.108.127204} {\bibfield  {journal}
  {\bibinfo  {journal} {Phys. Rev. Lett.}\ }\textbf {\bibinfo {volume} {108}},\
  \bibinfo {pages} {127204} (\bibinfo {year} {2012})}\BibitemShut {NoStop}%
\bibitem [{\citenamefont {Winter}\ \emph {et~al.}(2017)\citenamefont {Winter}
  \emph {et~al.}}]{Winter2017}%
  \BibitemOpen
  \bibfield  {author} {\bibinfo {author} {\bibfnamefont {S.~M.}\ \bibnamefont
  {Winter}} \emph {et~al.},\ }\bibfield  {title} {\bibinfo {title} {Breakdown
  of magnons in a strongly spin-orbital coupled magnet},\ }\href
  {https://doi.org/10.1038/s41467-017-01177-0} {\bibfield  {journal} {\bibinfo
  {journal} {Nature Communications}\ }\textbf {\bibinfo {volume} {8}},\
  \bibinfo {pages} {1152} (\bibinfo {year} {2017})}\BibitemShut {NoStop}%
\bibitem [{\citenamefont {Ferris}\ and\ \citenamefont
  {Vidal}(2012)}]{Ferris2012}%
  \BibitemOpen
  \bibfield  {author} {\bibinfo {author} {\bibfnamefont {A.~J.}\ \bibnamefont
  {Ferris}}\ and\ \bibinfo {author} {\bibfnamefont {G.}~\bibnamefont {Vidal}},\
  }\bibfield  {title} {\bibinfo {title} {Perfect sampling with unitary tensor
  networks},\ }\href {https://doi.org/10.1103/PhysRevB.85.165146} {\bibfield
  {journal} {\bibinfo  {journal} {Phys. Rev. B}\ }\textbf {\bibinfo {volume}
  {85}},\ \bibinfo {pages} {165146} (\bibinfo {year} {2012})}\BibitemShut
  {NoStop}%
\bibitem [{\citenamefont {Jin}\ \emph {et~al.}(2023)\citenamefont {Jin},
  \citenamefont {Kadow}, \citenamefont {Knap},\ and\ \citenamefont
  {Knolle}}]{Hui-Ke2023}%
  \BibitemOpen
  \bibfield  {author} {\bibinfo {author} {\bibfnamefont {H.-K.}\ \bibnamefont
  {Jin}}, \bibinfo {author} {\bibfnamefont {W.}~\bibnamefont {Kadow}}, \bibinfo
  {author} {\bibfnamefont {M.}~\bibnamefont {Knap}},\ and\ \bibinfo {author}
  {\bibfnamefont {J.}~\bibnamefont {Knolle}},\ }\bibfield  {title} {\bibinfo
  {title} {Kinetic {F}erromagnetism and {T}opological {M}agnons of the
  {Hole-Doped} {K}itaev {S}pin {L}iquid},\ }\href
  {https://arxiv.org/abs/2309.15153} {\bibfield  {journal} {\bibinfo  {journal}
  {arXiv:2309.15153}\ } (\bibinfo {year} {2023})}\BibitemShut {NoStop}%
\bibitem [{\citenamefont {White}\ and\ \citenamefont
  {Affleck}(2001)}]{White2001}%
  \BibitemOpen
  \bibfield  {author} {\bibinfo {author} {\bibfnamefont {S.~R.}\ \bibnamefont
  {White}}\ and\ \bibinfo {author} {\bibfnamefont {I.}~\bibnamefont
  {Affleck}},\ }\bibfield  {title} {\bibinfo {title} {Density matrix
  renormalization group analysis of the {Nagaoka} polaron in the
  two-dimensional $t\ensuremath{-}{J}$ model},\ }\href
  {https://doi.org/10.1103/PhysRevB.64.024411} {\bibfield  {journal} {\bibinfo
  {journal} {Phys. Rev. B}\ }\textbf {\bibinfo {volume} {64}},\ \bibinfo
  {pages} {024411} (\bibinfo {year} {2001})}\BibitemShut {NoStop}%
\bibitem [{\citenamefont {Bohrdt}\ \emph {et~al.}(2018)\citenamefont {Bohrdt},
  \citenamefont {Greif}, \citenamefont {Demler}, \citenamefont {Knap},\ and\
  \citenamefont {Grusdt}}]{Bohrdt2018}%
  \BibitemOpen
  \bibfield  {author} {\bibinfo {author} {\bibfnamefont {A.}~\bibnamefont
  {Bohrdt}}, \bibinfo {author} {\bibfnamefont {D.}~\bibnamefont {Greif}},
  \bibinfo {author} {\bibfnamefont {E.}~\bibnamefont {Demler}}, \bibinfo
  {author} {\bibfnamefont {M.}~\bibnamefont {Knap}},\ and\ \bibinfo {author}
  {\bibfnamefont {F.}~\bibnamefont {Grusdt}},\ }\bibfield  {title} {\bibinfo
  {title} {Angle-resolved photoemission spectroscopy with quantum gas
  microscopes},\ }\href {https://doi.org/10.1103/PhysRevB.97.125117} {\bibfield
   {journal} {\bibinfo  {journal} {Phys. Rev. B}\ }\textbf {\bibinfo {volume}
  {97}},\ \bibinfo {pages} {125117} (\bibinfo {year} {2018})}\BibitemShut
  {NoStop}%
\bibitem [{\citenamefont {Hwan~Chun}\ \emph {et~al.}(2015)\citenamefont
  {Hwan~Chun} \emph {et~al.}}]{HwanChun2015}%
  \BibitemOpen
  \bibfield  {author} {\bibinfo {author} {\bibfnamefont {S.}~\bibnamefont
  {Hwan~Chun}} \emph {et~al.},\ }\bibfield  {title} {\bibinfo {title} {Direct
  evidence for dominant bond-directional interactions in a honeycomb lattice
  iridate {Na}$_2${IrO}$_3$},\ }\href {https://doi.org/10.1038/nphys3322}
  {\bibfield  {journal} {\bibinfo  {journal} {Nature Physics}\ }\textbf
  {\bibinfo {volume} {11}},\ \bibinfo {pages} {462} (\bibinfo {year}
  {2015})}\BibitemShut {NoStop}%
\bibitem [{\citenamefont {Lin}\ \emph {et~al.}(2021)\citenamefont {Lin},
  \citenamefont {Moreschini},\ and\ \citenamefont {Lanzara}}]{Lin2021}%
  \BibitemOpen
  \bibfield  {author} {\bibinfo {author} {\bibfnamefont {C.-Y.}\ \bibnamefont
  {Lin}}, \bibinfo {author} {\bibfnamefont {L.}~\bibnamefont {Moreschini}},\
  and\ \bibinfo {author} {\bibfnamefont {A.}~\bibnamefont {Lanzara}},\
  }\bibfield  {title} {\bibinfo {title} {Present and future trends in spin
  {ARPES}},\ }\href {https://doi.org/10.1209/0295-5075/ac0c87} {\bibfield
  {journal} {\bibinfo  {journal} {Europhysics Letters}\ }\textbf {\bibinfo
  {volume} {134}},\ \bibinfo {pages} {57001} (\bibinfo {year}
  {2021})}\BibitemShut {NoStop}%
\bibitem [{\citenamefont {Motome}\ \emph {et~al.}(2020)\citenamefont {Motome},
  \citenamefont {Sano}, \citenamefont {Jang}, \citenamefont {Sugita},\ and\
  \citenamefont {Kato}}]{Motome2020}%
  \BibitemOpen
  \bibfield  {author} {\bibinfo {author} {\bibfnamefont {Y.}~\bibnamefont
  {Motome}}, \bibinfo {author} {\bibfnamefont {R.}~\bibnamefont {Sano}},
  \bibinfo {author} {\bibfnamefont {S.}~\bibnamefont {Jang}}, \bibinfo {author}
  {\bibfnamefont {Y.}~\bibnamefont {Sugita}},\ and\ \bibinfo {author}
  {\bibfnamefont {Y.}~\bibnamefont {Kato}},\ }\bibfield  {title} {\bibinfo
  {title} {Materials design of {Kitaev} spin liquids beyond the
  {Jackeli}–{Khaliullin} mechanism},\ }\href
  {https://doi.org/10.1088/1361-648X/ab8525} {\bibfield  {journal} {\bibinfo
  {journal} {Journal of Physics: Condensed Matter}\ }\textbf {\bibinfo {volume}
  {32}},\ \bibinfo {pages} {404001} (\bibinfo {year} {2020})}\BibitemShut
  {NoStop}%
\bibitem [{\citenamefont {Knolle}\ \emph {et~al.}(2014)\citenamefont {Knolle},
  \citenamefont {Kovrizhin}, \citenamefont {Chalker},\ and\ \citenamefont
  {Moessner}}]{Knolle2014}%
  \BibitemOpen
  \bibfield  {author} {\bibinfo {author} {\bibfnamefont {J.}~\bibnamefont
  {Knolle}}, \bibinfo {author} {\bibfnamefont {D.~L.}\ \bibnamefont
  {Kovrizhin}}, \bibinfo {author} {\bibfnamefont {J.~T.}\ \bibnamefont
  {Chalker}},\ and\ \bibinfo {author} {\bibfnamefont {R.}~\bibnamefont
  {Moessner}},\ }\bibfield  {title} {\bibinfo {title} {Dynamics of a
  {Two}-{Dimensional} {Quantum} {Spin} {Liquid}: {Signatures} of {Emergent}
  {Majorana} {Fermions} and {Fluxes}},\ }\href
  {https://doi.org/10.1103/PhysRevLett.112.207203} {\bibfield  {journal}
  {\bibinfo  {journal} {Phys. Rev. Lett.}\ }\textbf {\bibinfo {volume} {112}},\
  \bibinfo {pages} {207203} (\bibinfo {year} {2014})}\BibitemShut {NoStop}%
\bibitem [{\citenamefont {Feldmeier}\ \emph {et~al.}(2020)\citenamefont
  {Feldmeier}, \citenamefont {Natori}, \citenamefont {Knap},\ and\
  \citenamefont {Knolle}}]{Feldmeier2020}%
  \BibitemOpen
  \bibfield  {author} {\bibinfo {author} {\bibfnamefont {J.}~\bibnamefont
  {Feldmeier}}, \bibinfo {author} {\bibfnamefont {W.}~\bibnamefont {Natori}},
  \bibinfo {author} {\bibfnamefont {M.}~\bibnamefont {Knap}},\ and\ \bibinfo
  {author} {\bibfnamefont {J.}~\bibnamefont {Knolle}},\ }\bibfield  {title}
  {\bibinfo {title} {Local probes for charge-neutral edge states in
  two-dimensional quantum magnets},\ }\href
  {https://doi.org/10.1103/PhysRevB.102.134423} {\bibfield  {journal} {\bibinfo
   {journal} {Phys. Rev. B}\ }\textbf {\bibinfo {volume} {102}},\ \bibinfo
  {pages} {134423} (\bibinfo {year} {2020})}\BibitemShut {NoStop}%
\bibitem [{\citenamefont {König}\ \emph {et~al.}(2020)\citenamefont {König},
  \citenamefont {Randeria},\ and\ \citenamefont {Jäck}}]{Koenig2020}%
  \BibitemOpen
  \bibfield  {author} {\bibinfo {author} {\bibfnamefont {E.~J.}\ \bibnamefont
  {König}}, \bibinfo {author} {\bibfnamefont {M.~T.}\ \bibnamefont
  {Randeria}},\ and\ \bibinfo {author} {\bibfnamefont {B.}~\bibnamefont
  {Jäck}},\ }\bibfield  {title} {\bibinfo {title} {Tunneling {Spectroscopy} of
  {Quantum} {Spin} {Liquids}},\ }\href
  {https://doi.org/10.1103/PhysRevLett.125.267206} {\bibfield  {journal}
  {\bibinfo  {journal} {Phys. Rev. Lett.}\ }\textbf {\bibinfo {volume} {125}},\
  \bibinfo {pages} {267206} (\bibinfo {year} {2020})}\BibitemShut {NoStop}%
\bibitem [{\citenamefont {Kao}\ \emph {et~al.}(2023)\citenamefont {Kao},
  \citenamefont {Perkins},\ and\ \citenamefont {Halász}}]{Kao2023}%
  \BibitemOpen
  \bibfield  {author} {\bibinfo {author} {\bibfnamefont {W.-H.}\ \bibnamefont
  {Kao}}, \bibinfo {author} {\bibfnamefont {N.~B.}\ \bibnamefont {Perkins}},\
  and\ \bibinfo {author} {\bibfnamefont {G.~B.}\ \bibnamefont {Halász}},\
  }\bibfield  {title} {\bibinfo {title} {Vacancy spectroscopy of non-{Abelian}
  {Kitaev} spin liquids},\ }\href {https://arxiv.org/abs/2307.10376} {\bibfield
   {journal} {\bibinfo  {journal} {arXiv:2307.10376}\ } (\bibinfo {year}
  {2023})}\BibitemShut {NoStop}%
\bibitem [{\citenamefont {Peng}\ \emph {et~al.}(2021)\citenamefont {Peng},
  \citenamefont {Jiang}, \citenamefont {Devereaux},\ and\ \citenamefont
  {Jiang}}]{Peng2021}%
  \BibitemOpen
  \bibfield  {author} {\bibinfo {author} {\bibfnamefont {C.}~\bibnamefont
  {Peng}}, \bibinfo {author} {\bibfnamefont {Y.-F.}\ \bibnamefont {Jiang}},
  \bibinfo {author} {\bibfnamefont {T.~P.}\ \bibnamefont {Devereaux}},\ and\
  \bibinfo {author} {\bibfnamefont {H.-C.}\ \bibnamefont {Jiang}},\ }\bibfield
  {title} {\bibinfo {title} {Precursor of pair-density wave in doping {Kitaev}
  spin liquid on the honeycomb lattice},\ }\href
  {https://doi.org/10.1038/s41535-021-00363-0} {\bibfield  {journal} {\bibinfo
  {journal} {npj Quantum Materials}\ }\textbf {\bibinfo {volume} {6}},\
  \bibinfo {pages} {1} (\bibinfo {year} {2021})}\BibitemShut {NoStop}%
\bibitem [{\citenamefont {Shitade}\ \emph {et~al.}(2009)\citenamefont
  {Shitade}, \citenamefont {Katsura}, \citenamefont
  {Kune\ifmmode~\check{s}\else \v{s}\fi{}}, \citenamefont {Qi}, \citenamefont
  {Zhang},\ and\ \citenamefont {Nagaosa}}]{Shitade2009}%
  \BibitemOpen
  \bibfield  {author} {\bibinfo {author} {\bibfnamefont {A.}~\bibnamefont
  {Shitade}}, \bibinfo {author} {\bibfnamefont {H.}~\bibnamefont {Katsura}},
  \bibinfo {author} {\bibfnamefont {J.}~\bibnamefont
  {Kune\ifmmode~\check{s}\else \v{s}\fi{}}}, \bibinfo {author} {\bibfnamefont
  {X.-L.}\ \bibnamefont {Qi}}, \bibinfo {author} {\bibfnamefont {S.-C.}\
  \bibnamefont {Zhang}},\ and\ \bibinfo {author} {\bibfnamefont
  {N.}~\bibnamefont {Nagaosa}},\ }\bibfield  {title} {\bibinfo {title} {Quantum
  spin hall effect in a transition metal oxide
  $\mathrm{{Na}}_{2}\mathrm{{IrO}}_{3}$},\ }\href
  {https://doi.org/10.1103/PhysRevLett.102.256403} {\bibfield  {journal}
  {\bibinfo  {journal} {Phys. Rev. Lett.}\ }\textbf {\bibinfo {volume} {102}},\
  \bibinfo {pages} {256403} (\bibinfo {year} {2009})}\BibitemShut {NoStop}%
\bibitem [{\citenamefont {Hauschild}\ and\ \citenamefont
  {Pollmann}(2018)}]{Hauschild2018}%
  \BibitemOpen
  \bibfield  {author} {\bibinfo {author} {\bibfnamefont {J.}~\bibnamefont
  {Hauschild}}\ and\ \bibinfo {author} {\bibfnamefont {F.}~\bibnamefont
  {Pollmann}},\ }\bibfield  {title} {\bibinfo {title} {{Efficient numerical
  simulations with Tensor Networks: Tensor Network Python (TeNPy)}},\ }\href
  {https://doi.org/10.21468/SciPostPhysLectNotes.5} {\bibfield  {journal}
  {\bibinfo  {journal} {SciPost Phys. Lect. Notes}\ ,\ \bibinfo {pages} {5}}
  (\bibinfo {year} {2018})},\ \bibinfo {note} {code available from
  \url{https://github.com/tenpy/tenpy}}\BibitemShut {NoStop}%
\bibitem [{\citenamefont {Lieb}(1994)}]{Lieb1994}%
  \BibitemOpen
  \bibfield  {author} {\bibinfo {author} {\bibfnamefont {E.~H.}\ \bibnamefont
  {Lieb}},\ }\bibfield  {title} {\bibinfo {title} {Flux phase of the
  half-filled band},\ }\href {https://doi.org/10.1103/PhysRevLett.73.2158}
  {\bibfield  {journal} {\bibinfo  {journal} {Phys. Rev. Lett.}\ }\textbf
  {\bibinfo {volume} {73}},\ \bibinfo {pages} {2158} (\bibinfo {year}
  {1994})}\BibitemShut {NoStop}%
\bibitem [{\citenamefont {Zaletel}\ \emph {et~al.}(2015)\citenamefont
  {Zaletel}, \citenamefont {Mong}, \citenamefont {Karrasch}, \citenamefont
  {Moore},\ and\ \citenamefont {Pollmann}}]{Zaletel2014}%
  \BibitemOpen
  \bibfield  {author} {\bibinfo {author} {\bibfnamefont {M.~P.}\ \bibnamefont
  {Zaletel}}, \bibinfo {author} {\bibfnamefont {R.~S.~K.}\ \bibnamefont
  {Mong}}, \bibinfo {author} {\bibfnamefont {C.}~\bibnamefont {Karrasch}},
  \bibinfo {author} {\bibfnamefont {J.~E.}\ \bibnamefont {Moore}},\ and\
  \bibinfo {author} {\bibfnamefont {F.}~\bibnamefont {Pollmann}},\ }\bibfield
  {title} {\bibinfo {title} {Time-evolving a matrix product state with
  long-ranged interactions},\ }\href
  {https://doi.org/10.1103/PhysRevB.91.165112} {\bibfield  {journal} {\bibinfo
  {journal} {Phys. Rev. B}\ }\textbf {\bibinfo {volume} {91}},\ \bibinfo
  {pages} {165112} (\bibinfo {year} {2015})}\BibitemShut {NoStop}%
\bibitem [{\citenamefont {Paeckel}\ \emph {et~al.}(2019)\citenamefont {Paeckel}
  \emph {et~al.}}]{Paeckel2019}%
  \BibitemOpen
  \bibfield  {author} {\bibinfo {author} {\bibfnamefont {S.}~\bibnamefont
  {Paeckel}} \emph {et~al.},\ }\bibfield  {title} {\bibinfo {title}
  {Time-evolution methods for matrix-product states},\ }\href
  {https://doi.org/https://doi.org/10.1016/j.aop.2019.167998} {\bibfield
  {journal} {\bibinfo  {journal} {Annals of Physics}\ }\textbf {\bibinfo
  {volume} {411}},\ \bibinfo {pages} {167998} (\bibinfo {year}
  {2019})}\BibitemShut {NoStop}%
\bibitem [{\citenamefont {Luscher}\ \emph {et~al.}(2006)\citenamefont
  {Luscher}, \citenamefont {Laeuchli}, \citenamefont {Zheng},\ and\
  \citenamefont {Sushkov}}]{Luscher2006}%
  \BibitemOpen
  \bibfield  {author} {\bibinfo {author} {\bibfnamefont {A.}~\bibnamefont
  {Luscher}}, \bibinfo {author} {\bibfnamefont {A.}~\bibnamefont {Laeuchli}},
  \bibinfo {author} {\bibfnamefont {W.}~\bibnamefont {Zheng}},\ and\ \bibinfo
  {author} {\bibfnamefont {O.~P.}\ \bibnamefont {Sushkov}},\ }\bibfield
  {title} {\bibinfo {title} {Single-hole properties of the t-{J} model on the
  honeycomb lattice},\ }\href {https://doi.org/10.1103/PhysRevB.73.155118}
  {\bibfield  {journal} {\bibinfo  {journal} {Phys. Rev. B}\ }\textbf {\bibinfo
  {volume} {73}},\ \bibinfo {pages} {155118} (\bibinfo {year}
  {2006})}\BibitemShut {NoStop}%
\bibitem [{\citenamefont {White}\ and\ \citenamefont
  {Affleck}(2008)}]{White2008}%
  \BibitemOpen
  \bibfield  {author} {\bibinfo {author} {\bibfnamefont {S.~R.}\ \bibnamefont
  {White}}\ and\ \bibinfo {author} {\bibfnamefont {I.}~\bibnamefont
  {Affleck}},\ }\bibfield  {title} {\bibinfo {title} {{S}pectral function for
  the ${S}=1$ {H}eisenberg antiferromagnetic chain},\ }\href
  {https://doi.org/10.1103/PhysRevB.77.134437} {\bibfield  {journal} {\bibinfo
  {journal} {Phys. Rev. B}\ }\textbf {\bibinfo {volume} {77}},\ \bibinfo
  {pages} {134437} (\bibinfo {year} {2008})}\BibitemShut {NoStop}%
\bibitem [{\citenamefont {Burnell}\ and\ \citenamefont
  {Nayak}(2011)}]{Burnell2011}%
  \BibitemOpen
  \bibfield  {author} {\bibinfo {author} {\bibfnamefont {F.~J.}\ \bibnamefont
  {Burnell}}\ and\ \bibinfo {author} {\bibfnamefont {C.}~\bibnamefont
  {Nayak}},\ }\bibfield  {title} {\bibinfo {title} {{SU}(2) slave fermion
  solution of the {Kitaev} honeycomb lattice model},\ }\href
  {https://doi.org/10.1103/PhysRevB.84.125125} {\bibfield  {journal} {\bibinfo
  {journal} {Phys. Rev. B}\ }\textbf {\bibinfo {volume} {84}},\ \bibinfo
  {pages} {125125} (\bibinfo {year} {2011})}\BibitemShut {NoStop}%
\bibitem [{\citenamefont {Kadow}\ \emph {et~al.}(2023)\citenamefont {Kadow},
  \citenamefont {Jin}, \citenamefont {Knolle},\ and\ \citenamefont
  {Knap}}]{Zenodo}%
  \BibitemOpen
  \bibfield  {author} {\bibinfo {author} {\bibfnamefont {W.}~\bibnamefont
  {Kadow}}, \bibinfo {author} {\bibfnamefont {H.-K.}\ \bibnamefont {Jin}},
  \bibinfo {author} {\bibfnamefont {J.}~\bibnamefont {Knolle}},\ and\ \bibinfo
  {author} {\bibfnamefont {M.}~\bibnamefont {Knap}},\ }\href
  {https://doi.org/10.5281/zenodo.8363616} {\bibinfo {title} {{Zenodo entry
  for: Single-hole spectra of Kitaev spin liquids: From dynamical Nagaoka
  ferromagnetism to spin-hole fractionalization}}} (\bibinfo {year}
  {2023})\BibitemShut {NoStop}%
\end{thebibliography}%

\bigskip

\noindent \textbf{\large Acknowledgements}

\noindent We thank Wonjune Choi for insightful discussions.
We acknowledge support from the Deutsche Forschungsgemeinschaft (DFG, German Research Foundation) under Germany's Excellence Strategy--EXC--2111--390814868, DFG grants No. KN1254/1-2, KN1254/2-1, and TRR 360 - 492547816 and from the European Research Council (ERC) under the European Unions Horizon 2020 research and innovation programme (Grant Agreement No. 851161), Imperial-TUM flagship partnership, as well as the Munich Quantum Valley, which is supported by the Bavarian state government with funds from the Hightech Agenda Bayern Plus.
We thank the Nanosystems Initiative Munich (NIM) funded by the German Excellence Initiative and the Leibniz Supercomputing Centre for access to their computational resources.\\

\noindent \textbf{\large Author contributions}

\noindent W.K. performed the numerical and analytical calculations and evaluated the data. H-K.J. calculated the orbital structure of the spectral function. The research was devised by J. K. and M.K. All authors contributed to analyzing the data, discussions, and the writing of the manuscript. \\

\noindent \textbf{\large Competing interests}

\noindent The authors declare no competing interests. \\

\end{document}